%% file: simuJCAP.tex
\documentclass[a4paper,11pt]{article}
\pdfoutput=1 % if your are submitting a pdflatex (i.e. if you have
\usepackage{jcappub} % for details on the use of the package, please
\usepackage[varg]{txfonts}
\usepackage[utf8]{inputenc}

\usepackage{multirow}

\usepackage{subfigure}
\usepackage{graphicx}
\usepackage{rotating}
\usepackage{longtable}
\graphicspath{ {./plots/} }

\usepackage[load-configurations=astronomy]{siunitx}
\DeclareSIUnit\parsec{pc}
\DeclareSIUnit\hubble{\ensuremath{h}}
\newcommand\ion[2]{#1$\,${\scshape{#2}}}%                       % ion, i.e., CII = \ion{C}{ii}
\newcommand{\sun}{\ensuremath{\odot}}%

\title{\bf New approach for precise computation of Lyman-alpha forest power spectrum with hydrodynamical simulations }

\author[a]{Arnaud Borde,}
\author[a,b]{Nathalie Palanque-Delabrouille,}
\author[a,f]{Graziano Rossi,}
\author[c,d]{Matteo Viel,}
\author[e]{James S. Bolton,}
\author[a]{Christophe Yèche,}
\author[a]{Jean-Marc LeGoff,}
\author[a]{and Jim Rich}

\emailAdd{nathalie.palanque-delbrouille@cea.fr}
\emailAdd{arnaud.borde@cea.fr}

\affiliation[a]{CEA, Centre de Saclay, IRFU/SPP,  F-91191 Gif-sur-Yvette, France}
\affiliation[b]{Lawrence Berkeley National Laboratory, Berkeley, CA 94720, USA}
\affiliation[c]{INAF, Osservatorio Astronomico di Trieste, Via G. B. Tiepolo 11, 34131 Trieste, Italy}
\affiliation[d]{INFN/National Institute for Nuclear Physics, Via Valerio 2, I-34127 Trieste, Italy}
\affiliation[e]{School of Physics and Astronomy, University of Nottingham, University Park, Nottingham, NG7 2RD, UK}
\affiliation[f]{Department of Astronomy and Space Science, Sejong University, Seoul, 143-747, Korea}

\date{Received xx; accepted xx}

\abstract{Current experiments  are providing measurements of the flux power spectrum from  the Lyman-$\alpha$ forests observed in quasar spectra with unprecedented accuracy. Their interpretation  in terms of cosmological constraints requires specific simulations of at least equivalent precision. 
In this paper, we present  a suite of cosmological $N$-body simulations with cold dark matter and baryons, specifically aiming at modeling the low-density regions of the inter-galactic medium as probed by the Lyman-$\alpha$ forests at high redshift. The simulations were run using the \texttt{GADGET-3} code and  were designed to match the requirements imposed by the quality of the current SDSS-III/BOSS or forthcoming SDSS-IV/eBOSS data. They are  made using either $2 \times 768^3 \simeq 1$~billion or $2 \times 192^3 \simeq 14$~million particles, spanning volumes ranging from $(25 \,{\rm Mpc.h^{-1}})^3$ for high-resolution simulations to $(100 \,{\rm Mpc.h^{-1}})^3$ for large-volume ones.  Using a splicing technique, the resolution is further enhanced to reach the equivalent of simulations with $2 \times 3072^3 \simeq 58$~billion particles in a $(100 \,{\rm Mpc.h^{-1}} )^3$ box size, i.e. a  mean mass per gas particle of  $1.2\times10^{5}M_\sun . h^{-1}$. We show that the resulting power spectrum is accurate at the 2\% level over the full range  from a few Mpc to several tens of Mpc.  
We explore the effect on the one-dimensional transmitted-flux power spectrum of four cosmological parameters ($n_s$, $\sigma_8$, $\Omega_m$ and $H_0$) and two astrophysical parameters ($T_0$ and $\gamma$) that are related to the heating rate of the intergalactic medium. By varying the input parameters around a central model chosen to be in agreement with the latest Planck results, we built a grid of simulations that allows the study of the impact on the flux power spectrum of these six relevant parameters.  
We improve upon previous studies by not only measuring the effect of each  parameter individually, but also probing the impact of the simultaneous variation of each pair of  parameters. We thus provide a full second-order expansion, including cross-terms, around our central model. We check the validity of the second-order expansion
with independent simulations obtained either with different cosmological parameters or different seeds. Finally, a comparison to the one-dimensional Lyman-$\alpha$ forest power spectrum obtained with BOSS by  \citet{Palanque-Delabrouille2013} shows an excellent agreement.}

\begin{document}
\maketitle
\flushbottom

\section{Introduction}\label{sec:intro}
\input{intro}

%%%%%%%%%%

\section{Simulation grid}\label{sec:grid}
\input{grid}

%%%%%%%%%%

\section{Pipeline}\label{sec:pipeline}
\input{pipeline}

%%%%%%%%%%

\section{Convergence tests}\label{sec:tests}
\input{tests}

%%%%%%%%%%

\section{Splicing}\label{sec:splicing}
\input{splicing}

%%%%%%%%%%

\section{Results and discussions}\label{sec:results}
\input{results}

%%%%%%%%%%

\section{Conclusions}\label{sec:conclusion}
\input{conclusion}

%%%%%%%%%%

\appendix
\section{List of all simulations}\label{sec:appendix}
\input{appendix}

\acknowledgments

We acknowledge PRACE (Partnership for Advanced Computing in Europe) for awarding us access to resource curie-thin nodes based in France at TGCC, under allocation number 2012071264.\\
This work was also granted access to the resources of CCRT under the allocation 2013-t2013047004 made by
GENCI (Grand Equipement National de Calcul Intensif).\\
A.B., N.P.-D., G.R. and Ch.Y.  acknowledge  support from grant ANR-11-JS04-011-01 of Agence Nationale de la Recherche.\\
M.V. is supported by ERC-StG "CosmoIGM".\\
JSB acknowledges the support of a Royal Society University 
Research Fellowship.\\
We thank Volker Springel for making \texttt{GADGET-3} available to our team.

%%%%%%%%%%

\bibliographystyle{unsrtnat_arxiv}
\bibliography{biblio}

\end{document}

%% file: intro.tex
In the intergalactic medium, light is absorbed at the Lyman-$\alpha$ absorption wavelength
$\lambda_{\rm Ly\alpha} \sim \SI{1216}{\angstrom}$ by neutral hydrogen. Combined with cosmological redshifting,
it produces an absorption spectrum which is observed on any background source as a map of transmission fraction
as a function of redshift \citep{Lynds1971}. For light sources at sufficiently high redshift for the absorption
of the intergalactic matter to be sufficiently strong, the continuous nature of the absorption spectrum is easily observable
as the Lyman-$\alpha$ forest. Although this spectrum can be seen as a series of merged absorption lines, simulations
have shown that it is in reality a map of density fluctuations in the intervening intergalactic medium seen in
redshift space, with peaks of absorption at the density peaks of the absorbing gas \citep{Bi1992, Miralda-Escude1993}.
Moreover, the fluctuations in the Lyman-$\alpha$ forest absorption can be used as a tracer of the varying density of
intergalactic gas expected from the growth of structure from primordial fluctuations in the Universe \citep{Croft1998}.
An intergalactic medium heated exclusively by photo-ionization can be modeled with hydrodynamic simulations
\citep{Cen1992a, Zhang1995, Hernquist1996, Hui1997, Hui1997a, Viel2004} and the physics at play in this model is well understood.
%However, mechanisms such as radiative transfer effects during hydrogen and helium reionisation \citep{Abel1999} or the mechanical effects of galactic winds and quasar outflows can quickly complicate this simple picture, making simulations even more useful.

The amplitude and shape of the power spectrum of mass fluctuations can be measured through the information embedded in the
Lyman-$\alpha$ forest  observable  in quasar spectra \citep{Croft1998, Gnedin1998, Hui1999, Gaztanaga1999, Nusser1999, Feng2000, McDonald2000, Hui2001}.
These can later be used to constrain cosmology \citep{Alcock1979, Hui1999, McDonald1999, Croft2002}, the baryonic acoustic
oscillation peak position \citep{McDonald2007} or the sum of the masses of neutrinos \citep{Seljak2005, Viel2010}.
Small numbers of high-resolution spectra were first used to measure the Lyman-$\alpha$ forest power spectrum: 1 Keck HIRES
spectrum \citep{Croft1998}, 19 spectra from the Hershel telescope on La Palma or the AAT \citep{Croft1999}, 8 Keck HIRES spectra
\citep{McDonald2000}, a set of 30 Keck HIRES and 23 Keck LRIS spectra \citep{Croft2002}, or a set of 27 high resolution UVES/VLT
QSO spectra at redshifts $\sim$ 2 to 3 \citep{Kim2004, Kim2004a, Viel2004}. The Sloan Digital Sky Survey \citep{York2000} lead to
a major breakthrough providing a much larger sample of 3035 medium-resolution ($R = \lambda / \Delta \lambda_{\rm FWHM} \approx 2000$) quasar
spectra for the measurement of the Lyman-$\alpha$ forest power spectrum by \citet{McDonald2006}. The large number of observed
quasars allowed detailed measurements with well characterized errors of the power spectrum up to larger scales, probing the
linear regime and providing cosmological constraints \citep{McDonald2005, Seljak2005}.

The next step is carried out by the  Sloan Digital Sky Survey III \citep{Eisenstein2011} through the Baryon Oscillation
Spectroscopic Survey (BOSS, \citet{Dawson2012}). Quasars at redshift $z>2$, which are  useful for  Lyman-$\alpha$ forest analyses,
are specifically targeted, leading to a much higher number of such quasar spectra than in previous surveys (\citet{Dawson2012}
and references therein). It  thus allows a measurement of the Lyman-$\alpha$ power spectrum in both three-dimensional and
one-dimensional redshift space. The 60,000 quasars spectra with Lyman-$\alpha$ forest absorption \citep{Paris2012, Lee2013} of the
Data Release 9 \citep{Ahn2012} have already permitted the measurement of the BAO peak position and new constraints on the
history of the expansion of the universe \citep{Busca2012, Slosar2013, Kirkby2013} using the three-dimensional power spectrum.
A measurement of the one-dimensional power spectrum $P_{1D}$ with a significant improvement over previous studies in the achieved precision has also been conducted
\citep{Palanque-Delabrouille2013}.  Other background sources, such as Lyman-break galaxies, are also being investigated for a dense mapping of the Lyman-$\alpha$ forest \citep{Lee2013a}.

Whereas the measurement of the three-dimensional power spectrum uses only information from the flux correlation of pixel pairs
in different quasar spectra and thus provides information on rather large scales, the one-dimensional power spectrum$P_{1D}$, defined by
\begin{equation}
  P_{1D}(k_\parallel) = \int_0^\infty \frac{dk_\perp k_\perp}{2\pi}\, P_{3D}(k_\parallel,k_\perp) ~,
\label{eq:P3DvsP1D}
\end{equation}
 uses the correlation
of pixel pairs on the same quasar spectrum and thus provides a complementary, useful information on smaller scales that are fundamental to constrain the physical parameters of the Lyman-$\alpha$ forest. The one-dimensional $P_{1D}$ is probing scales  at the transition
from linear to non-linear regime. Therefore, cosmological simulations are required to provide insight on the non-linear physics of the intergalactic medium on the small scales  probed by $P_{1D}$. Such simulations are then used to constrain various
cosmological and astrophysical parameters that have an effect on the power spectrum \citep{Viel2004, McDonald2005, Viel2005,
Bolton2008, Viel2010, Bird2011, Bird2011a}.

Here, we present a set of 28 cosmological smoothed particles hydrodynamics (SPH) and $N$-body simulations that reproduce the impact
on the one-dimensional matter power spectrum of the values taken by the most relevant cosmological and astrophysical parameters.
Only the baryonic particles undergoes a SPH treatment, i.e. they receive an additional hydrodynamic acceleration, and their internal
entropy per unit mass is evolved as an independent thermodynamic variable. All our simulations are run using \texttt{GADGET-3}, last described by \cite{Springel2005}. The requirements in terms of box size, resolution and redshift coverage of our simulations
were derived from the Data Release 9 quasar catalogue \citep{Paris2012, Ahn2012} of the Baryon Oscillation
Spectroscopic Survey \citep{Dawson2012}. We  extrapolate these requirements so that this suite of simulations may also be used for
future spectroscopic surveys such as eBOSS\footnote{\url{http://www.sdss3.org/future/eboss.php} and
\url{http://www.sdss3.org/future/sdss4.pdf}} (planned for 2014-2018) or DESI\footnote{\url{http://desi.lbl.gov}}
\citep{Schlegel2011} (2018-2023). The full suite of simulations will be made available upon request to the authors.

The outline of the paper is as follows. In section~\ref{sec:grid} we describe our grid and the values chosen for the different
parameters we varied. In section~\ref{sec:pipeline}, we present the simulations pipeline, along with our solutions to issues such as the generation of the initial conditions or 
the  radiative cooling and heating processes that occur in the intergalactic medium (IGM). In section~\ref{sec:tests} we present tests that
were made to determine the required characteristics of our simulations in the light of our goals. We  describe, in section~\ref{sec:splicing}, the splicing
technique we apply in order to obtain simulations with the desired resolution and box size. We demonstrate the validity  of our grid approach and present final discussions on this suite of simulations
 in section~\ref{sec:results}. Conclusions and perspectives are given in section~\ref{sec:conclusion}. A recapitulation of all the simulations performed for this study is given in appendix~\ref{sec:appendix}.

%% file: grid.tex
Ideally, in order to derive confidence intervals on each parameter of a cosmological model with eight to ten
free parameters, one would like to compute theoretical predictions for thousands of models, exploring most of the
parameter space. 
Statistical frameworks have been studied to optimize the precision of the model for a reduced number of simulations, such as  Latin hypercube sampling \citep{Tang1993}. While this method is superior to
a random sampling of the parameters for instance as regards the attained precision \citep{Mckay2000}, it still requires a large number of simulations.  Latin hypercube sampling has only been tested so
far to predict the power spectrum on large-scales, using low resolution simulations, of order $128^3$ particles for a 450 ${\rm Mpc}.h^{-1}$ box \citep{Heitmann2009, Schneider2011}.

When dealing with Lyman-$\alpha$ data, running large numbers of simulations is not possible due to the high execution time of each
hydrodynamical simulation. Hence, various approximate methods have been 
developed in which a restricted number
of simulations is used either to calibrate a flux-to-matter power-spectrum bias function or to Taylor expand the
flux power-spectrum with respect to cosmological parameters in the vicinity of a best-fit model. For cosmological predictions of the power spectrum in the Lyman-alpha regime where hydrodynamical
simulations are required, the grid approach as presented in \citep{Viel2006} is  generally adopted (cf. \citep{Wang2013} for instance for a recent application). This is the method we have selected for this work. 

\subsection{Methodology}
\begin{figure}
\centering
\includegraphics[width=.5\linewidth]{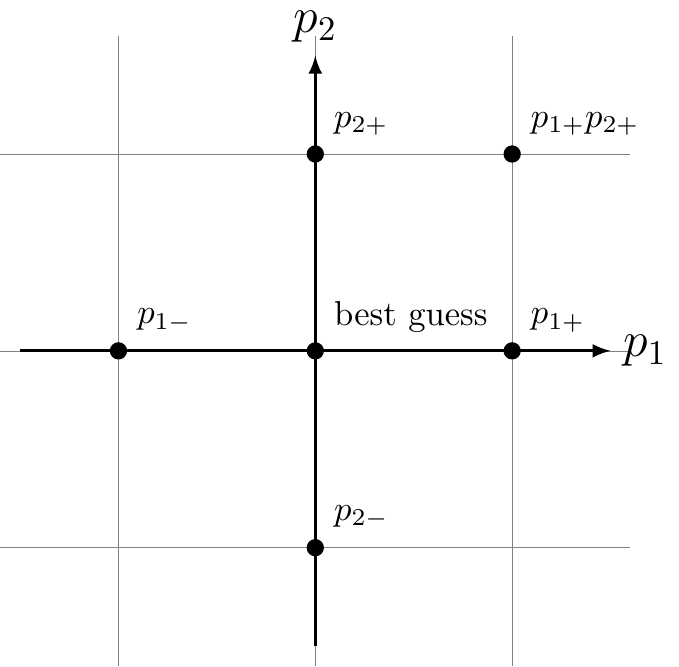}
\caption{Illustration of the required grid for a second-order Taylor expansion in a two-dimensional parameter space.}\label{fig:grid}
\end{figure}

We model the variations of the one-dimensional power spectrum with relevant input parameters by a second-order Taylor expansion about our best-guess  model:
\begin{eqnarray}\label{eq:taylor}
f(\mathbf{x} + \mathbf{\Delta x}) &=& f(\mathbf{x})\nonumber \\
&&+ \sum\limits_{i} \frac{\partial f}{\partial x_i} (\mathbf{x}) \Delta x_i  \nonumber \\
&& + \frac{1}{2} \sum\limits_{i} \sum\limits_{j} \frac{\partial^2 f}{\partial x_i \partial x_j}(\mathbf{x}) \Delta x_i \Delta x_j
\,.
\end{eqnarray}
With $n$ parameters, the total number of simulations required to get the Taylor expansion coefficients 
is  $1 + 2n + n(n-1)/2$, where the terms account for, respectively, the central (or best-guess) model, two other values  of each parameter to derive the  first and second-order
derivatives, and the simultaneous variation of each pair of parameters to compute the cross derivatives (cf. figure~\ref{fig:grid}). 
With this lattice, all derivatives are approximated  to second order except the
cross derivatives which are approximated to first order. This approximation is justified by the fact that the parameters are reasonably decoupled, and it allowed us to reduce
the CPU time consumption since second-order cross derivatives would require additional $n(n-1)/2$ simulations.

\subsection{Simulation variable parameters}
To model of the physics of the Universe, we  introduced two categories of parameters that are varied
in the simulations: cosmological parameters that describe the cosmological model in the simplest case of $\Lambda$CDM assuming
a flat Universe with mass-less neutrinos, and astrophysical parameters that model the astrophysics within the IGM and the relation
between temperature  and  density of the gas. 
A summary of all simulations performed to compute the coefficients of the Taylor expansion is given in appendix~\ref{sec:appendix}.

\subsubsection{ Cosmological parameters}

This first category contains four parameters: 
the amplitude of the matter power spectrum $\sigma_8$, the spectral
index of primordial density fluctuations $n_s$, the matter density $\Omega_m$ and the Hubble constant $H_0$.
The values  for our central model are in agreement with the latest best-fit
values from Planck \citep{PlanckCollaboration2013}, which we recall in table~\ref{table:planck_best_fit}.
\begin{table}[h]
\begin{center}
\begin{tabular}{|lcc|}
\hline
Parameter & Best fit & 68\% limits\\
\hline
$n_s$\dotfill & $0.9624$ & $0.9616 \pm 0.0094$\\
$\sigma_8$\dotfill & $0.8344$ & $0.834 \pm 0.027$\\
$\Omega_m$\dotfill & $0.3175$ & $0.314 \pm 0.020$\\
$H_0$\dotfill & $67.11$ & $67.4 \pm 1.4$\\
\hline
\end{tabular}
\end{center}
\caption{Cosmological parameters values from Planck temperature power spectrum alone. We give best fit and 68\%
confidence limits.}\label{table:planck_best_fit}
\end{table}
We chose the range of variation for these parameters so as to include other recent constraints from the Wilkinson Microwave
Anisotropy Probe seven years data \citep{Komatsu2011}, the South Pole Telescope data \citep{Hou2012} and the SuperNova Legacy Survey
three year data \citep{Conley2011, Sullivan2011}, thus taking into account the fact that results from Planck for $H_0$ (respectively
$\Omega_m$) are low (respectively high) compared to other measurements. Central values at  redshift $z=0$ and range for each of the cosmological parameters are given in table~\ref{table:grid_parameters}.  We also give in the appendix the values of $\sigma_8$ at redshift $z=3.0$, (pivot redshift of Ly$\alpha$ power spectrum measurements of \cite{Palanque-Delabrouille2013} and \cite{McDonald2005}), and of the shape parameter $\Gamma=\Omega_m h$, often used  in the first Ly$\alpha$  studies.

\subsubsection{ Astrophysical parameters}

This second category includes two redshift-dependent parameters that describe the temperature-density relation of the IGM  for $\rho / \left< \rho \right> \leq 10$:
\begin{equation}\label{eq:t_rho}
T(\rho, z) = T_0(z) \times \left( \frac{\rho}{\left< \rho \right>} \right)^{\gamma(z) - 1}\,,
\end{equation}
where $\rho$ is the baryonic density,  $T_0(z)$ is a normalization temperature  and  $\gamma(z)$ a logarithmic slope.
At the post processing step (cf. \ref{sec:Post-processing}), we  scaled the effective optical depth
$\tau_{\rm eff} = - \ln \left( \left< F \right> \right) = -\ln \left( \left< e^{-\tau} \right> \right)$, where $F$ is the flux and $\tau$
is the optical depth,  so that it followed a power law $\tau_{\rm eff}(z) =  \tau_A\times (1+z) ^{\tau_{S}}$. We allow for different  mean flux  normalizations and evolutions with redshift by varying the   parameters $\tau_A$ and $\tau_S$.

In the absence of a clear consensus on the heating history of the IGM, we took the $T(\rho)$ measurements from \citet{Becker2011}
assuming $\gamma = 1.3$ as our central model, and we chose a wide variation around these values so that other recent measurements
\citep{Garzilli2012, Lidz2010, Schaye2000} fall into the explored range.
The evolution with redshift of $\gamma(z)$ and $T_0(z)$ in our simulations is therefore designed to reproduce the $T(\rho)$ measurements presented by
\citet{Becker2011} through an adaptation of the cooling routines in the simulation
code. Thus we only need to fix those two parameters at a given redshift, in our case $z=3.0$, which corresponds to the  central redshift
of our study. In practice, we do not set $T_0(z=3)$ and $\gamma (z=3)$ but instead use two internal code parameters, \texttt{AMPL} and 
\texttt{GRAD}, that alter the amplitude and  density dependence of the photo-ionization heating rates, such that
$\epsilon_f = \mathtt{AMPL} \times \delta^{\mathtt{GRAD}} \times \epsilon_i$ where $\epsilon$'s are the heating rates and $\delta$ is the
over-density. $T_0$ and $\gamma$ are evaluated after the simulations have run, as explained at the end of section \ref{sec:pipeline}.
Given the one-to-one correspondence between ($T_0$, $\gamma$) and (\texttt{AMPL}, 
\texttt{GRAD}), we prefer to keep on quoting  $T_0$ and $\gamma$  since these parameters have a physical meaning and can be compared to other studies.

 There are also a number of additional astrophysical effects which will impact 
on the Lyman-alpha flux power spectrum which we have not considered in this 
work. For example, the hydrogen reionisation history will alter the pressure 
smoothing scale of gas in the IGM, particularly at redshifts approaching the 
tail-end of the reionisation at $z\sim6$ \citep{Gnedin1998b}. 
Galactic winds will impact on the distribution of HI around haloes, although 
the filling factor of the winds will be small at $z>2$ \citep{Theuns2002}.  Finally helium reionisation may induce fluctuations in the 
ionisation and thermal state of the IGM at $z\sim3$, impacting on the power 
spectrum at large scales \citep{McQuinn2011}. Note, 
however, these will generally influence the power spectrum to a lesser  extent 
than the variations in the effective optical depth and the thermal state of the 
IGM we consider here, see e.g. \citep{McDonald2005b}.  As a 
result, we do not include them within our current analysis.  These second 
order effects will nevertheless be important to consider for precision 
measurements with the Lyman-alpha forest power spectrum, although note that 
modelling these self-consistently will still ultimately require radiation 
hydrodynamics calculations which are currently computationally prohibitive. 

\subsubsection{Grid values}
The central values and variation ranges of the parameters of our study are summarized In table \ref{table:grid_parameters}.
With six varying parameters, this represents a total of 28 cosmological simulations in our grid. We also varied the mean flux as explained in the previous section, but this was done a posteriori and did not require any additional simulation. 

\begin{table}[h]
\begin{center}
\begin{tabular}{|lll|}
\hline
Parameter & Central value & Range\\
\hline
$n_s$\dotfill & $0.96$ & $\pm\,0.05$\\
$\sigma_8$\dotfill & $0.83$ & $\pm\,0.05$\\
$\Omega_m$\dotfill & $0.31$ & $\pm\,0.05$\\
$H_0$\dotfill & $67.5$ & $\pm\,5$\\
$T_0(z=3)$\dotfill & $14000$ & $\pm\,7000$\\
$\gamma(z=3)$\dotfill & $1.3$ & $\pm\,0.3$\\
$\tau_A$ \dotfill & 0.0025 & $\pm\,0.0020$\\
$\tau_S$ \dotfill & 3.7& $\pm\,0.4$\\
\hline
\end{tabular}
\end{center}
\caption{Central values and variation ranges of the cosmological parameters  for our simulation grid.}\label{table:grid_parameters}
\end{table}

%% file: pipeline.tex
\begin{figure*}
\centering
\includegraphics[width=\textwidth]{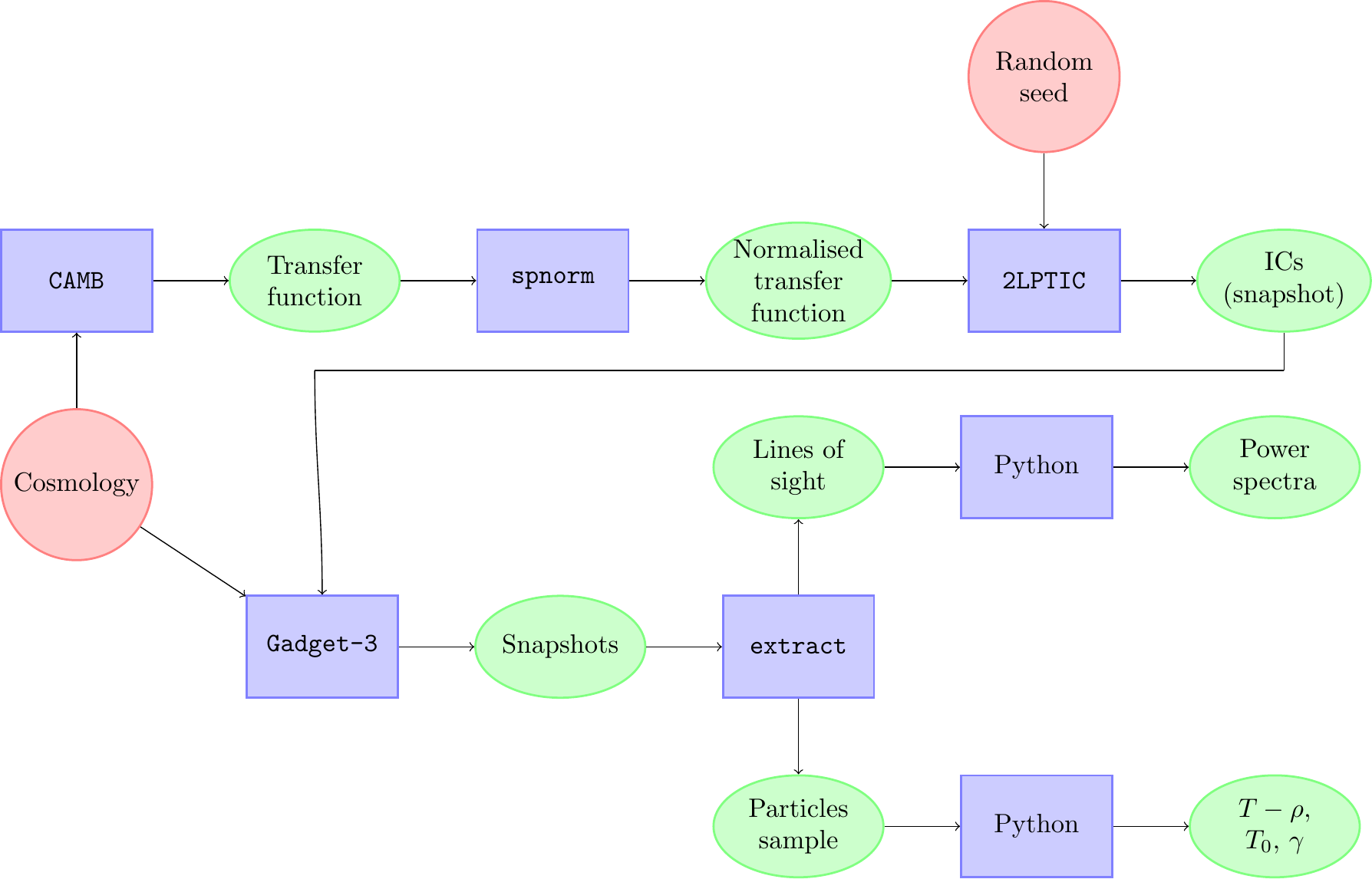}
\caption{Our simulation pipeline: red circles represent input from the user, blue rectangles are software packages and scripts and yellow
ellipses correspond to outputs from the 
software.}\label{fig:pipeline}
\end{figure*}

All the components of our simulation  work flow are represented on figure \ref{fig:pipeline}.
The first part of the pipeline is the production of the initial condition snapshot. This is done in the linear approximation with
perturbations treated up to  second order. The simulations are then performed using both $N$-body and hydrodynamic (SPH) treatments.
The post-processing stage takes the result of the simulations and computes the power spectra that will be compared to  data through the
Taylor expansion described earlier.

The products of our suite of simulations are obtained at 13 predefined redshifts, equally spaced every $\Delta z = 0.2$ from $z=2.2$
to 4.6. Our selection of redshifts reflects the possibilities of current and forthcoming large-scale spectroscopic surveys. In SDSS, the lower bound
results from the UV cut-off of CCDs  at $\lambda\sim 350$~nm that prevents the observation of Lyman-$\alpha$ below $z\sim2.2$. The upper
bound results from the quasar luminosity function that peaks near $z\sim2$ and drops significantly at $z>3$. The density of QSOs at
$z>4$  is of order 0.3 per square degree to a limiting magnitude $g<22$ as is the case for SDSS-III/BOSS, and even to $g<23$ as expected for the future DESI survey, only reaches a density of 2 per square degree. This is less than an
order of magnitude smaller than at $z\sim 2$~\citep{Palanque-Delabrouille2013a}.

\subsection{\ttfamily CAMB}
The Code for Anisotropies in the Microwave Background (\texttt{CAMB})\footnote{\url{http://camb.info}} \citep{Lewis2000} is a
numerical Boltzmann code written in Fortran 90. It is a parallelized line-of-sight integration code developed from \texttt{CMBFAST}
\citep{Seljak1996} and \texttt{COSMICS} \citep{Bertschinger1995}, which is widely used (and thus tested) to calculate not only
the lensed cosmic microwave background temperature and polarization spectra but also linear matter power spectra for different species
of particles (in our case baryons and dark matter). 

CAMB is here used to compute the transfer functions and linear power spectra that
will be used in the next step to compute the initial displacement of particles.

\subsection{\ttfamily 2LPT}
All our simulations are tuned to obtained a given $\sigma_8$ at $z=0$. This is done  with the \texttt{spnorm} Python script that  rescales the total matter power spectrum $P_S^{\rm CAMB}$ issued from CAMB before  generating the initial conditions such that 
\begin{equation}
P_S = P_S^{\rm CAMB} \times \left[ \frac{\sigma_8(z_i)}{\sigma_8^{\rm CAMB}(z_i)}
\right]^2 \, ,
\end{equation}
where $z_i$ is the redshift at which the initial conditions are run, $\sigma_8^{\rm CAMB}$ is the value of $\sigma_8$ obtained with CAMB  for a chosen  cosmological model, and 
\begin{equation}
\sigma_8(z_i) = \frac{\sigma_8 (z=0)}{\sigma_8^{\rm CAMB} (z=0)}\times \sigma_8^{\rm CAMB}(z_i) \,.
\end{equation}
 Thus, in the 2LPTIC code, the power spectra are taken from CAMB at $z=0$ and scaled back to the initial redshift $z_{\rm i}=30$ by explicitly forcing the simulation to achieve the desired value of $\sigma_8$  at $z=0$. While \texttt{CAMB} includes radiation, this is not the case for \texttt{GADGET}. 
However, the impact of the radiation component is very small in terms of the 
matter power spectrum at the scales relevant for the present work.  
Consequently, the low-redshift evolution of the simulation reproduces the 
matter and transmitted flux power spectra in a regime in which the radiation 
contribution can be safely ignored.
The rescaled power spectra are then used as input to the \texttt{2LPTIC}\footnote{\url{http://cosmo.nyu.edu/roman/2LPT/}} code that provides initial conditions based on second-order Lagrangian
Perturbation Theory (2LPT), rather than first-order (Zel'dovich approximation). The choice of second-order precision initial
conditions is motived by the discussion in \citet{Crocce2006} and the fact that we also run cosmological simulations including neutrinos as a new particle type~\citep{Rossi2014}. Indeed, because of their high velocity,   neutrinos  require initial conditions taken at rather low redshift in order 
to reduce  Poisson noise  \citep{Ali-Haimoud2012a, Bird2011a}.
Initial conditions for all the grid simulations are run with the same seed.
%A comparison of the two methods at redshift $30$ is given in figure \ref{fig:za_2lpt}.

%\begin{figure}
%\centering
%\includegraphics[width=\linewidth]{dummy.png}
%\caption{comparison ZA/2LPT}
%\label{fig:za_2lpt}
%\end{figure}

\subsection{\ttfamily Gadget-3}
\texttt{GADGET-3} (GAlaxies with Dark matter and Gas intEracT) is a massively parallel tree-SPH code for 
cosmological simulations, originally developed by Volker Springel and collaborators \citep{Springel2001,Springel2005}.
It is written in ANSI C, and uses the standardized message passing interface (\texttt{MPI}) along with several
open-source libraries (\texttt{GSL}\footnote{\url{http://www.gnu.org/software/gsl/}},
\texttt{FFTW}\footnote{\url{http://www.fftw.org/}}). Gravitational interactions are computed via a hierarchical multipole
expansion using the standard $N$-body method, and gas-dynamics are followed with smoothed particle
hydrodynamics (SPH); collisionless dark matter and gas are both represented by particles. 

Since its original version (\texttt{GADGET-1}), the code underwent a series of improvements and 
optimizations over several years (\texttt{GADGET-2} and \texttt{3}), to maximize the work-load balance and the efficiency
in memory consumption and communication bandwidth. In what follows, we briefly describe the key features of the code.

\begin{figure*}
\centering
\subfigure[Baryonic gas]{
\centering
\includegraphics[height=0.87\textheight]{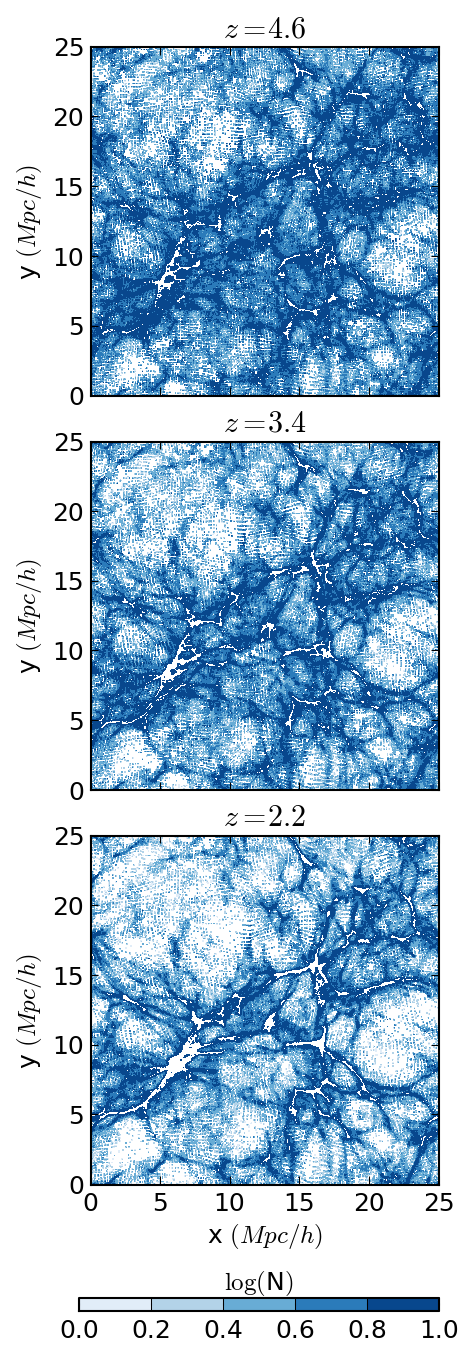}
}
\hfill
\subfigure[Dark matter]{
\centering
\includegraphics[height=0.87\textheight]{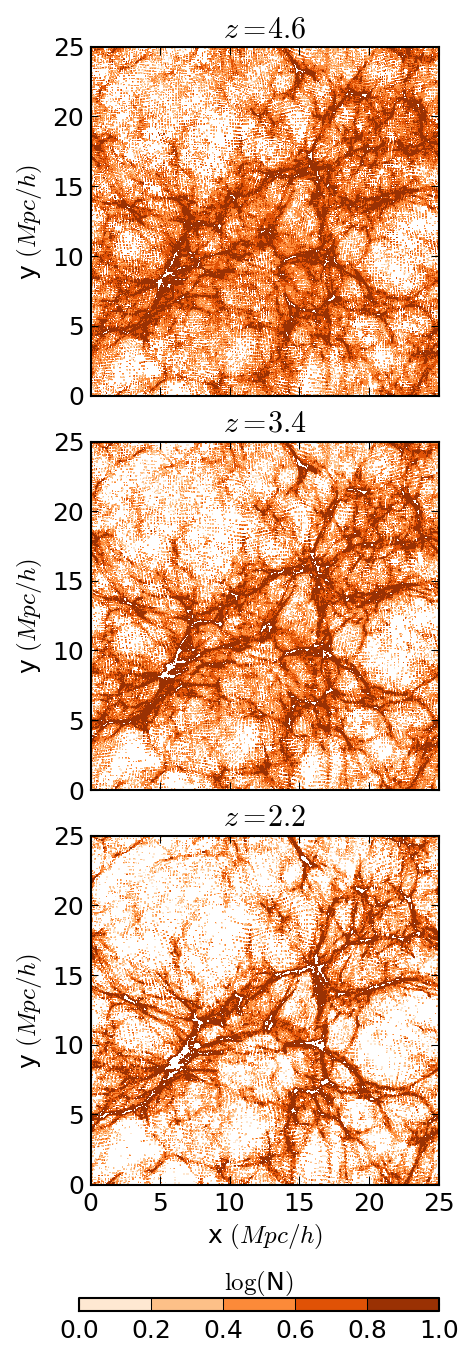}
}
\caption{Slice of baryon and dark matter snapshots ($\SI{2.5}{\mega\parsec\per\hubble}$ depth), at three different redshifts, extracted from a simulation with $192^3$ particles per type in a 
$(\SI{25}{\mega\parsec\per\hubble})^3$ box. As expected, there are very few
differences between the distributions for the two types of particles. Color represents particle number density. Out-of-scale densities (whether underflow or overflow) are white.}
\label{fig:slices}
\end{figure*}

\texttt{GADGET-3} follows a collisionless fluid with the standard $N$-body method, and an ideal gas with smoothed
particle hydrodynamics (SPH). The code solves simultaneously for the dynamics of the collisionless component
and of the ideal gas, both subject to and coupled by gravity in an expanding background space. The $N$-body
implementation only differs from other cosmological codes by the accuracy of the gravitational field computation.
A number of further physical processes have also been implemented in \texttt{GADGET-3}, from radiative cooling/heating
physics to non-standard dark matter dynamics, star formation and feedback. In figure \ref{fig:slices}, we present the
evolution of a filament with redshift and in figure \ref{fig:snapshot} we show the image of a snapshot made with \texttt{splotch}
\footnote{\url{http://www.mpa-garching.mpg.de/~kdolag/Splotch}}. Such realizations can be used for visual confirmation before
quantitative analysis as well as for public outreach and education.

Several optimization strategies have been added in \texttt{GADGET-3}. These include a Peano-Hilbert space decomposition,
a massively parallel version of the Fast Fourier Transform library,  the possibility of splitting the simulation
data across several files (to facilitate and speed-up the input/output process), and the fact that the code can
be run on an arbitrary number of processors. In its current version, \texttt{GADGET-3} is highly efficient in memory
consumption (it allocates up to $80$ bytes per particle) and communication bandwidth, is versatile and flexible,
accurate and fast. Another important aspect  is the scalability of the code, i.e. its
performance when the number of processors is increased, which has currently been tested up to {16,000}
cores.

We started all our simulations at $z=30$ with initial conditions based on second-order Lagrangian perturbation theory\citep{Crocce2006}, and adopted the same gravitational softening for the different species considered (i.e. gas, dark matter, stars), which however varies with the length of the box and the size of the mesh chosen. Specifically, we set the gravitational softening length  to 0.8~kpc$/h$ for the simulation having $25\,{\rm Mpc}/h$ boxsize and resolution $2\times768^3$, while the softening is 3.25~kpc$/h$ for the other two runs, i.e., the $25\,{\rm Mpc}/h$ boxsize and  $2\times 192^3$ resolution, and the $100\,{\rm Mpc}/h$ boxsize and  $2\times768^3$ resolution.  We used the `QUICKLYA' routine in \texttt{GADGET-3} to simulate the Lyman-$\alpha$ forest, assuming the gas of  primordial composition with a helium mass fraction of $Y=0.24$. We neglect metals and the evolution of elementary  abundances, as well as feedback processes and galactic winds. Along the lines of \citet{Viel2010}, we adopted a simplified criterion for star formation: all gas particles whose overdensity with respect to the mean is above 1000 and whose temperature is less than $10^5\,{\rm K}$ are turned into star particles immediately. 

\subsection{\ttfamily extract}
The \texttt{GADGET-3} snapshots contain various fields among which the position $\mathbf{p}$ and velocity $\mathbf{v}$ for both dark
matter and gas particles.  It also contains fields that are specific to the SPH treatment of gas particles: internal energy $U$,
density $\rho$, electron fraction ${N_e}$, hydrogen fraction $N_{\rm H}$ and smoothing length $h$. We use these fields to
extract two samples:
\begin{itemize}

\item{\bf a particle sample:} we extract a subsample of particles to study the temperature-density relation. For each particle the
temperature is derived with the formula 
\begin{equation}
k_B T = U \times (\gamma -1) \times \mu  M_{\rm H}\, ,
\end{equation}
where $\mu = 1/(X_{\rm H} (0.75+N_e) + 0.25)$. $\gamma$ is the adiabatic index ($5/3$ for monoatomic gas), $M_{\rm H}$ is the mass of an
hydrogen atom, $k_B$ is the Boltzmann constant and $X_{\rm H}$ is the hydrogen fraction by mass. Figure~\ref{fig:t-rho} illustrates typical temperature-density diagrams  obtained from this particle sample. \\

\item{\bf a line of sight sample:} {\bf following the traditional procedure in one-dimensional flux power studies~\citep{gnedin2002, Croft2002}}, we extract lines of sight (LOS) from the simulation cube choosing random origin and axis. For each
pixel of each LOS, we derive density $\rho$, temperature $T$, peculiar velocity $v$ and optical depth $\tau$, all for \ion{H}{I} only
using the SPH equation:
\begin{equation}
A(\mathbf{r}) = \sum\limits_{j} m_j \frac{A_j}{\rho_j} W\left( \left| \mathbf{r} - \mathbf{r}_j \right|, h_j \right)
\end{equation}
where $A$ is a scalar quantity, $\mathbf{r}$ a position in the cube, $h$ the smoothing length, and $W$ a kernel function. The
index $j$ runs on all particles. We use the 3D cubic spline kernel:\\
\begin{equation}
W(q_j) =
\begin{cases}
 [1 + q_j^2 (-1.5+0.75q_j)] . \frac{1}{\pi}& \left| q_j \right| \leq 1\\
[0.25(2-q_j)^3] . \frac{1}{\pi}& 1 < \left| q_j \right| \leq 2\\
0 & \left| q_j \right| \geq 2
\end{cases}
\end{equation}
%\begin{tabular}{rl}\\
%$W(q_j) = \left\{
%\begin{array}{ll}
%  [1 + q_j^2 (-1.5+0.75q_j)]. \frac{1}{\pi}& \left| q_j \right| \leq 1\\
%0.25.(2-q_j)^3 . \frac{1}{\pi}& 1 < \left| q_j \right| \leq 2 \\
%0 &  \left| q_j \right| \geq 2
%\end{array}
%\right. $\\
%\end{tabular}
%\\
where $q_j = \left|\mathbf{r} - \mathbf{r_j}\right| / h_j$.
These LOS are not mock spectra, in the sense that they do not match any properties (such as noise, resolution, metals absorption, \dots) of observational data. 
The quantity of  particular interest for our study  is the optical depth for \ion{H}{I}, from which we  compute the  transmitted flux for each pixel. 
\end{itemize}

\subsection{Post-processing}\label{sec:Post-processing}
The post-processing stage allows us to extract two categories of outputs. The first one is the large-box high-resolution power spectrum that is derived by an appropriate combination of the power spectra from 3 lower-resolution or smaller-box simulations, using
the splicing technique described in section \ref{sec:splicing}. 
At this stage, we  fix the photo-ionization rate (or equivalently the UV flux) by requiring the effective optical depth at each redshift to follow the empirical power law $\tau_{\rm eff}(z) =  \tau_A\times (1+z) ^{\tau_{S}}$, where $\tau_A=0.0025$ and $\tau_S=3.7$ in agreement with observations~\citep{Meiksin2009}. The rescaling coefficients,  determined  independently for each redshift bin using all the line-of-sight pixels, are typically between $-20\%$ and $+20\%$. We perform this normalization  a posteriori since it is computationally much cheaper than finding and fixing the appropriate photo-ionization rate a priori for each of the simulations.  As explained in \cite{Theuns2005}, however, this is justified by the fact that when the gas is highly ionized and in photo-ionization equilibrium, as is the case for the Lyman-$\alpha$ forest, the total heating rate per unit volume is independent of the amplitude of the UV flux. Gas dynamics can thus be considered not to be affected by the UV flux. The power spectrum is then computed from the scaled flux, and averaged over all  lines of sight.

The second category  results from the particle sample. It is used to derive the parameters $T_0(z)$ and $\gamma(z)$ in the IGM. This is performed by estimating the location of the most populated region of the diagram using the mode of the 2D distribution, for  particles lying in the region defined by 
$\log(\delta) \in [-0.5, 0.0]$ and $\log (T/\SI{1}{K}) < 5.0$, with $\delta = \rho / \left< \rho \right>$. Given the large tail of particles toward the high temperature regions where clusters of galaxies reside, in particular at low redshift, the mode  was preferred to  the mean since it is not affected by the precise choice of the $(\delta, T)$ bounds used to define the IGM. We estimate the mode 
by taking bins of $\SI{1000}{\kelvin}$ and computing the barycenter of the five highest bins. A linear fit is then performed
using these points.

%% file: tests.tex
We base our minimal requirements for the resolution and  box size   of our simulations on the  largest currently-available spectroscopic  survey: SDSS-III/BOSS \citep{Dawson2012}.
Those requirements  are  driven  by  the extension of the Lyman-$\alpha$ forest that can be probed experimentally and by the measurement errors on the power spectrum , which set the convergence levels to be achieved in the simulations.

The quasar coadded spectra provided by the SDSS pipeline \citep{Bolton2012}
are computed with a constant pixel width of $\Delta v = \SI{69}{\kilo\metre\per\second}$. The largest mode is bounded
by the Nyquist-Shannon limit at $k_{\rm Nyquist} = \pi / \Delta v = \SI{4.5e-2}{(\kilo\metre\per\second)^{-1}}$. Instrumental constraints, however, make this theoretical limit  very difficult to obtain with reasonable precision from data, and the largest mode measured in BOSS data by \citet{Palanque-Delabrouille2013} is $k_{\rm max} =  \SI{2.0e-2}{(\kilo\metre\per\second)^{-1}}$.
The smallest mode is driven by the extension  of the Lyman-$\alpha$ forest which lies between the Lyman-$\alpha$ and Lyman-$\beta$
emissions respectively at $\SI{1216}{\angstrom}$ and $\SI{1026}{\angstrom}$. The exploitable Lyman-$\alpha$ forest, however, is smaller than the separation of the two emission peaks due to their respective
widths. \citet{Palanque-Delabrouille2013} computed the 1D power spectrum from forest lengths corresponding to a third of the total available range in order to restrain the redshift span to $\Delta z=0.2$ at most. This led to $k_{\rm min} \sim \SI{1.0e-3}{(\kilo\metre\per\second)^{-1}}$.
We therefore consider simulations that should cover the minimal range $\SI{1e-3}{(\kilo\metre\per\second)^{-1}} < k < \SI{2e-2}{(\kilo\metre\per\second)^{-1}}$, 
which corresponds approximately to $\SI{0.1}{(\mega\parsec\per\hubble)^{-1}} < k < \SI{2}{(\mega\parsec\per\hubble)^{-1}}$ at $z\sim 3$.

In numerical simulations, the two relevant parameters  are the size of the box $L$ that  determines the smallest $k$-mode ($k_{\rm min} = 2\pi/L$), and
the ratio $N^{1/3}/L$, where $N$ is number of particles,  that drives the largest $k$-mode. One may note that due to the
computational algorithms used nowadays in simulations, such as smooth-particles hydrodynamics (SPH) or adaptive mesh
refinement (AMR) in which ``resolution follows density'', particle spacing in high-density regions  will be significantly smaller
than $L/N^{1/3}$. Because the 1D power spectrum results from an integral over the 3D power spectrum up to $k=\infty$ (cf. Eq.~\ref{eq:P3DvsP1D}), the resolution of the simulations has to be of the size of the smallest structures in the transverse direction. For structures  
in  local hydrostatic equilibrium, this would be the Jeans scale, of order a few 100~kpc at $z=3$.  In an SPH approach, over-dense regions  are sampled with much higher spatial resolution than average. Under-dense  regions, on the other hand, might not necessarily be in local hydrostatic equilibrium. 
The decisive solution to ensure that the simulations do resolve the relevant structures is therefore to perform  convergence tests.  

\begin{figure*}
\centering
\subfigure[Mass-resolution tests, the reference simulation has $L=\SI{20}{\mega\parsec\per\hubble}$ and $2\times1024^3$ particles.]{
  \centering
  \includegraphics[width=0.48\linewidth]{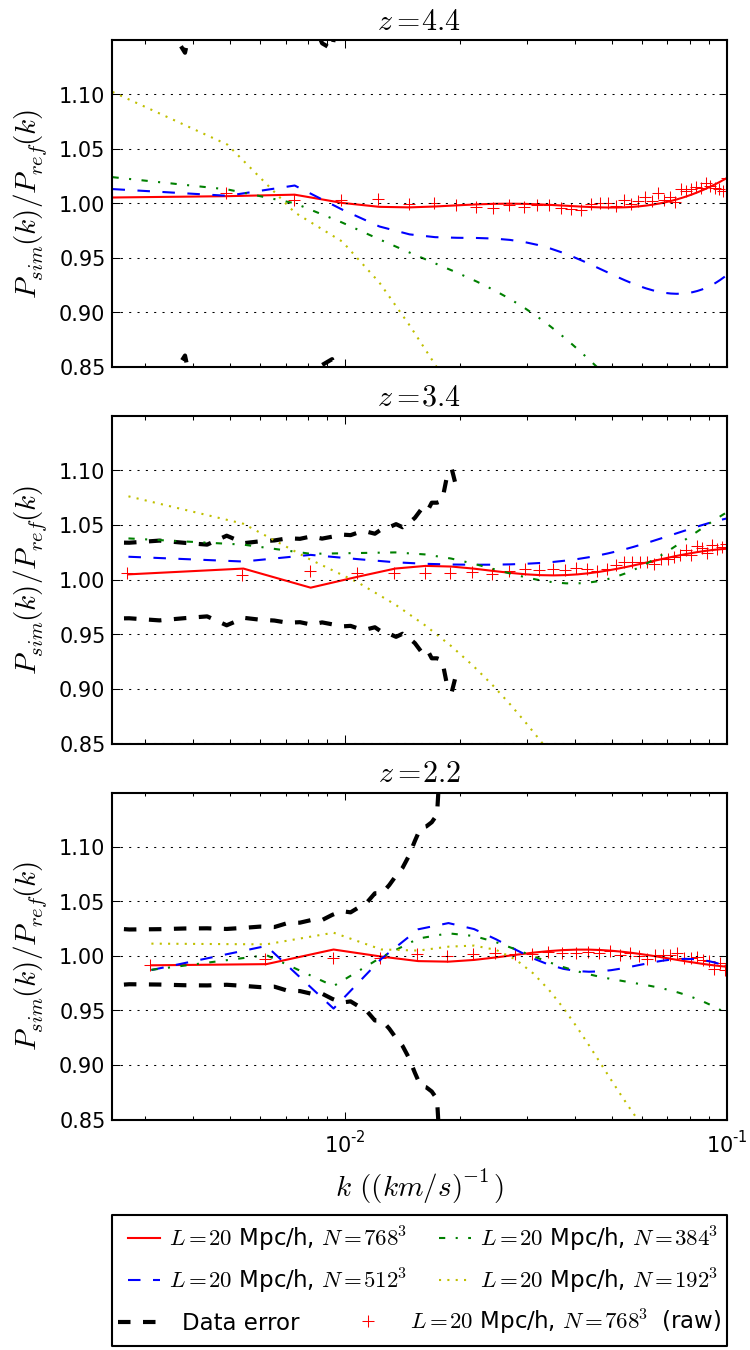}\label{fig:convergence_tests_resolution}
}
\hfill
\subfigure[Box-size tests, the reference simulation has $L=\SI{120}{\mega\parsec\per\hubble}$ and $2\times1024^3$ particles.]{
  \centering
  \includegraphics[width=0.48\linewidth]{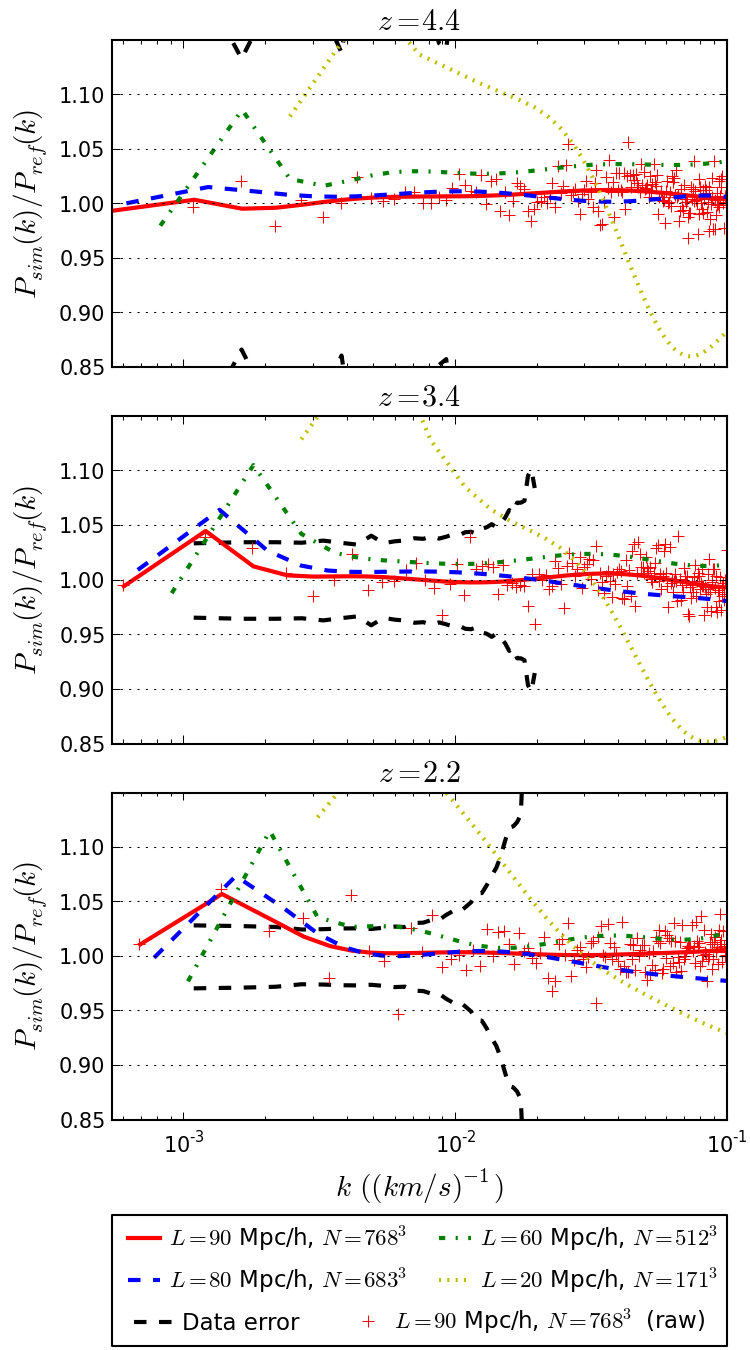}\label{fig:convergence_tests_boxsize}
}
\caption{Convergence tests for mass resolution and box size at three different redshifts. The curves are 5\textsuperscript{th}-order  polynomial functions fitted to
the data for better visibility. All values of the power spectrum ratio are shown  for the (20,768) (left) and the (90,768) (right) cases. The dashed black curves illustrate $1\sigma_{\rm stat}$ uncertainties measured in BOSS data. Data uncertainties exceed the plot boundaries at $z=4.2$.}\label{fig:convergence_tests}
\end{figure*}

The simulations  used for the convergence test are all run with the same random seed  and with the following cosmological parameters: $(\Omega_m, \Omega_bh^2, h, \sigma_8, n_s)= (0.31, 0.021, 0.675, 0.83, 0.96)$. 
We ran two sets of simulations: the first set  with simulations having the same box size $L$ of $\sim 20\,{\rm Mpc/}h$ but changing the particle loading $N^3$ and therefore the mass resolution, the second  with simulations having the same mass resolution but varying volumes, keeping $L/N$ fixed at $\sim 0.12$. These two sets are listed in table \ref{table:convergence_sets} and the results at three different redshifts are presented in figures
\ref{fig:convergence_tests_resolution} and \ref{fig:convergence_tests_boxsize}. Hereafter, we will use the notation ($L$,$N$) to represent a simulation with $N^3$ particles for each species (dark matter and baryons) in a box of size 
$L\:\SI{}{\mega\parsec\per\hubble}$ on a side. 

\begin{table}[h]
\begin{center}
\begin{tabular}{|c|c|}
\hline
Mass-resolution test & Box-size test\\
\hline
\textbf{(20,1024)} & \textbf{(120,1024)}\\
(20,768) & (90,768)\\
(20,512) & (80,683)\\
(20,384) & (60,512)\\
(20,192) & (20,171)\\
\hline
\end{tabular}
\end{center}
\caption{The two sets of simulations used for convergence tests, with the reference simulation indicated in bold.
$(L,N)$ refers to a simulation with $N^3$ particles per species in a box of size 
$L\:\SI{}{\mega\parsec\per\hubble}$ on a side.}\label{table:convergence_sets}
\end{table}

These convergence tests are  more stringent than what has been done before, justified by our aim to use our simulation suite for comparison to data of higher quality. For instance, to probe the effect of the box size, 
\cite{Viel2004} compared  to a reference simulation with $(L,N)=(120,200)$  and thus $L/N\sim 0.60$, and \cite{Bolton2009} to a reference simulation $(80,400)$  i.e.,  $L/N\sim 0.20$. This is to be compared to our $L/N$ of 0.12. As regards the convergence on the mass-resolution, we explored a similar range of mass-resolutions as \cite{Bolton2009} (in contrast, \cite{Viel2004} restricted to a minimum mass per particle 3 times larger), but using a $\SI{20}{\mega\parsec\per\hubble}$ box instead of $\SI{10}{\mega\parsec\per\hubble}$.

The most difficult redshifts at which to achieve convergence are  those at $z>3$, since the mean flux level becomes very small at such epochs and  under-dense regions, which are less well sampled than average in an SPH framework, are producing absorption. High redshift bins, however, are very important since gravitational collapse
tends to suppress the differences in the linear-theory power spectra. These bins therefore highlight primordial
differences between the matter power spectra resulting from different contributions of the various cosmological constituents. 
Some convergence problems can also arise at  $z\sim2$  due to the fact that strong systems, which are very non-linear, might  be simulated inadequately due to cosmic variance or lack of resolution. Low-redshift bins are also those where the measurements from QSO spectra have the smallest statistical error bars, making the convergence criteria tighter.

\subsection{Mass resolution}

We computed the ratio of the  power spectra of each of the simulations listed in the first column of table~\ref{table:convergence_sets} to the power spectrum of the (120,1024) simulation.
The results presented in figure \ref{fig:convergence_tests_resolution} show that  an excess of
power on large scales (small $k$) and a lack of power on small scales (large $k$) appear with decreasing resolution. As expected, this effect is stronger
at higher redshift where the Lyman-$\alpha$ forest  probes low density regions, 
which are less well resolved in the SPH treatment since it is the  mass (and not the spatial) resolution that is kept fixed. Further details about this effect
can be found in \citet{Bolton2009}. 

The dashed  curves in figure~\ref{fig:convergence_tests_resolution}   illustrate the level of the $1\sigma_{\rm stat}$  statistical uncertainties observed in the BOSS analysis  \citep{Palanque-Delabrouille2013} at each redshift. At $z=4.2$, the experimental uncertainties are larger than the maximum $\pm 15\%$ departure allowed on the plot and no longer appear. 

Simulations with a mass resolution at least as good as  for the (20,512) simulation all deviate by less than $2.5\%$ from the highest mass-resolution power spectrum over our minimal  $k$-range. This corresponds to a mean mass per particle of $M=2.2\times 10^5 M_\sun . h^{-1}$. Extending to $k_{\rm max}= \SI{0.1}{(\kilo\metre\per\second)^{-1}}$, the (20,512) simulation  deviates by $\sim 10\%$ at the largest redshift,.

%These two effects come mainly from the larger size of particles from lower resolution simulations: on the one hand,
%gravitational collapse tends to produce bigger clusters and on the other hand, small scale structures are not resolved.

\subsection{Box size}

The results of figure \ref{fig:convergence_tests_boxsize} show that the box size has an effect on all scales, and not only on the large
scales that approach  the Nyquist limit. This is due to the non-linear coupling of modes during gravitational evolution, and to the fact
that even on scales close to the box size, mass-fluctuations are not fully linear. 

As before, the dashed  curves in figure~\ref{fig:convergence_tests_boxsize} illustrate the level of the $1\sigma_{\rm stat}$  data uncertainties at each redshift. To reach $k_{\rm min} = \SI{1.0e-3}{(\kilo\metre\per\second)^{-1}}$, we see that we need a box size of at least $\SI{90} {\mega\parsec\per\hubble}$.  The most significant constraint comes  from the largest scales that cannot be probed (or not with adequate precision) otherwise. 

\subsection{Summary of convergence requirements}

In conclusion, the ideal simulation for our study should use a $\sim~\SI{100}{\mega\parsec\per\hubble}$ box and a mass resolution roughly equivalent to a (20,614) simulation, which translates into  $3072^3$ particles of each species. 
The  mean  mass of a gas particle is then $M=1.2\times10^{5}M_\sun . h^{-1}$. 

Although  convergence tests are specific to each problem and each statistical property for which convergence is sought, we can briefly compare to the results obtained by other studies. To infer the dark matter power spectrum from the Lyman-$\alpha$ forest in high-resolution QSO absorption spectra covering $0.003<k<\SI{0.03}{(\kilo\metre\per\second)^{-1}}$, \citet{Viel2004} chose a $(60,400)$ simulation, i.e. a mass per gas particle of $\sim 4\times 10^7 M_\sun . h^{-1}$.
To resolve the high redshift Lyman-$\alpha$ forest in smoothed particle hydrodynamics simulations,  a problem similar to our own, \citet{Bolton2009} found that a box size of at least $40 \,{\rm Mpc}.h^{-1}$ is preferable at all
redshifts. They also found that while a mean gas particle mass $M_{\rm gas}\leq1.6\times 10^6 M_\sun . h^{-1}$ is required at $z=2$, a mass resolution at least 8 times better is needed at $z=5$, i.e. $M_{\rm gas}\leq2\times 10^5 M_\sun . h^{-1}$. 
Our requirements are thus  more stringent than selected in past, both in terms of box size and mass resolution. 

Several tens of such simulations, as needed to compute our grid of cosmological simulations, 
would require several tens of millions of hours to be run, which is not an acceptable  computational time. We address  and solve this issue with the splicing technique presented in the next section.

%% file: splicing.tex
In the previous section we have estimated that simulating a flux power spectrum covering the range
$k = \SI{1e-3}{(\kilo\metre\per\second)^{-1}}$ to $k = \SI{2e-2}{(\kilo\metre\per\second)^{-1}}$) with a unique simulation
at sufficient precision for every redshift in the  range $2.2 < z < 4.6$ requires $N=3072^3$ particles of each species in a box of size $L=\SI{100}{\mega\parsec\per\hubble}$. To obtain  power spectra of equivalent resolution and box size in a reasonable computational time, we use the technique described in \cite{McDonald2003}. In this method, competing demands of large box 
size and high resolution are solved by splicing together the power spectra from pairs of large and small box simulations, using
$L=\SI{100}{\mega\parsec\per\hubble}$ for the large-scale power, and $L=\SI{25}{\mega\parsec\per\hubble}$ for the small-scale power,
both with $N=768^3$. One must then correct the large box size simulation for the lack of resolution, and the small box size for the
lack of non-linear coupling between the highest and the lowest $k$-modes. The corrections are computed using a transition $(25,192)$
simulation that has  same resolution as a $(100,768)$ and same box size as a $(25,768)$.

One needs to distinguish three regimes when computing the full power spectra:
\begin{itemize}
\item $\mathbf{k < k_{min, 25}}$, where $k_{min, 25} = 2\pi / \SI{25}{\mega\parsec\per\hubble}$ is the minimum $k$ present
in a $L=\SI{25}{\mega\parsec\per\hubble}$ box. The spliced flux power $P_{F}$ is the  flux power $P_{F,100,768}$ of the
$(100,768)$ simulation here taken as our reference, corrected for its low-resolution by a $k$-independent factor evaluated at $k_{min,25}$:
\[
P_{F} (k) = P_{F,100,768} (k) \times \frac{P_{F,25,768}(k_{min,25})}{P_{F,25,192}(k_{min,25})}\,.
\]
The possibility of using a constant factor for the largest $k$-modes has been tested in \cite{McDonald2003}.

\item $\mathbf{k_{min, 25} < k < k_{Nyq,100} /4}$, where $k_{Nyq,100} = 768\pi / \SI{100}{\mega\parsec\per\hubble}$ is the
Nyquist wave number of the large box. In this regime we use a similar correcting ratio, but taken at the wave number $k$ at which
the flux power is calculated:
\[
P_{F} (k) = P_{F,100,768} (k) \times \frac{P_{F,25,768}(k)}{P_{F,25,192}(k)}\,.
\]
This is mathematically equivalent to considering the high-resolution simulation (25,768) as our reference, and correcting it for its small box size.

\item $\mathbf{k > k_{Nyq,100}/4}$. At these large $k$-modes, the resolution correction is no longer a small factor. We
 thus take the $(25,768)$ simulation as our reference, and correct for its limited box size by a $k$-independent factor evaluated at the fixed 
splicing point $k = k_{Nyq,100}/4$:
\[
P_{F} (k) = P_{F,25,768} (k) \times \frac{P_{F,100,768}(k_{Nyq,100}/4)}{P_{F,25,192}(k_{Nyq,100}/4)}\,.
\]
\end{itemize}

\begin{figure*}
\centering
\includegraphics[width=\linewidth]{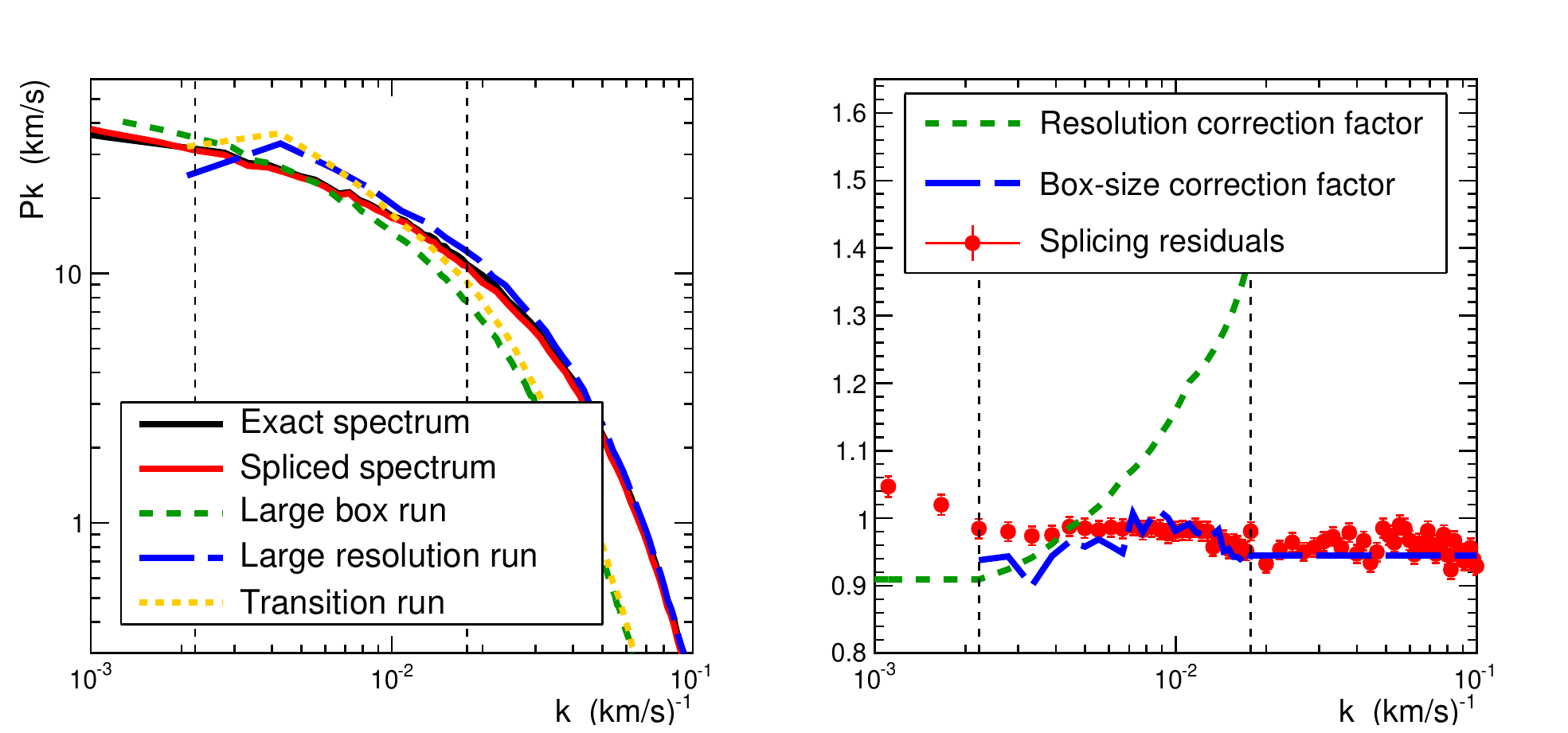}
\caption{Illustration of the splicing technique with simulations of either $64^3$ or $256^3$ particles in a volume of $(25\,{\rm Mpc}/h)^3$ or $(100\,{\rm Mpc}/h)^3$. Dashed vertical lines illustrate the regime boundaries. Left:  power spectrum at $z=3.0$ for the exact (100,1024) run, the spliced technique and the individual components entering the splicing. Right:  correction factors computed with respect to the power spectrum chosen as reference in each regime.  Although the correction presents a discontinuity, the spliced power spectrum is continuous by construction. Splicing residuals are overlaid.}
\label{fig:splicing}
\end{figure*}

The splicing technique is applied  for each  redshift at which we compute the power spectrum.  We illustrate the method and its accuracy on figure \ref{fig:splicing}, using a set of smaller-resolution  simulations to enhance the contrast between the different power spectra, as well as to allow the comparison to a full resolution run (labelled ``exact'' on the figure)  with 1024 particles in a $(100\, {\rm Mpc}/h)^3$ box.  For this illustration, the large box-size, the large resolution and the transition simulations are (100,256), (25,256) and (25,64) simulations respectively.  The spliced power spectrum obtained at $z=3.0$ is presented on the left of figure \ref{fig:splicing}, along with the exact power spectrum and the individual runs entering the splicing estimate. The  correction coefficient with respect to the reference power spectrum in each regime of $k$-modes is shown on the right plot of figure \ref{fig:splicing}.  In the intermediate regime, the resolution correction increases towards smaller scales, reaching 40\% for the set of simulations illustrated here. It shows less scale-dependence when taken as a box-size correction to the large-resolution power spectrum, but it is noisier since it requires taking the ratio of two simulations with different box sizes and thus different natural $k$-modes.   Although the correction factors show discontinuities at the boundary where the simulation chosen as reference changes, the spliced power spectrum is continuous by construction. In the large-mode splicing regime, at $k>k_{Nyq,100}/4$,  it is unclear whether a constant box-size correction or even any correction at all is indeed the optimal combination, since both the correction factor and the residuals are at the same level of about 0.95. This regime, however, is only probed by the medium resolution SDSS-III/BOSS data in the highest redshift bins where measurement uncertainties significantly exceed the splicing errors. Its optimization  is thus beyond the scope of this paper.

For the box size and resolution chosen for our simulation suite, the last regime begins at $k=5.3\times 10^{-2}\,(\rm km/s)^{-1}$ for $z=3.0$, which is beyond the maximum mode that can be reached with  BOSS or eBOSS data. The maximum
correction factor, obtained for $k=2.0\times 10^{-2}\,({\rm km/s})^{-1}$, is thus smaller than in the previous illustration. It ranges from 22\% at $z=4.6$ to 5\% at $z=2.2$.

\begin{figure}
\centering
\includegraphics[width=\linewidth]{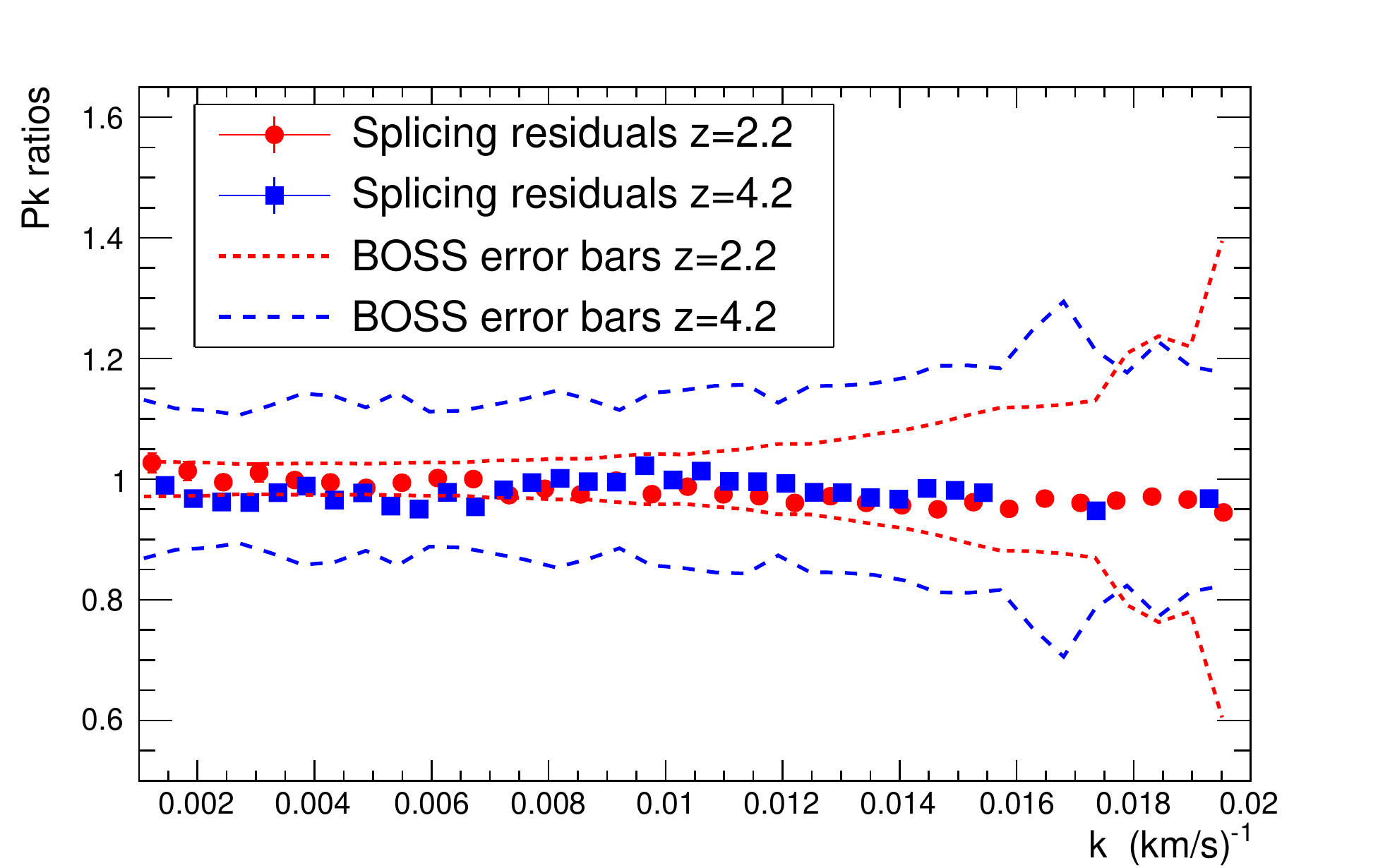}
\caption{Residuals of the  spliced to the exact power spectrum for $z=2.2$ and $z=4.2$. The dashed curves illustrates the level of statistical uncertainties in current data ($1\sigma$).}
\label{fig:splicing_residuals}
\end{figure}

We estimate the accuracy of the technique from the splicing residuals, defined as the ratio of the spliced to the exact power spectrum. The splicing residuals show no dependence with redshift. The  residuals at $z=3.0$ are overlaid on the right plot of figure \ref{fig:splicing}.  In  figure \ref{fig:splicing_residuals}, they  are plotted  for $z=2.2$ and $z=4.2$, along with  the  statistical uncertainty at the same redshifts obtained in the most recent BOSS analysis \citep{Palanque-Delabrouille2013}. Over the $k$-range of interest for BOSS data, the residuals have an average of $-0.98$ with an $rms$ of 0.01. The largest excess is seen near $k=10^{-3}\,({\rm Mpc}/h)^3$. A simulation with a larger box size would be needed to reduce the splicing residuals further. 
For the purpose of this study, the splicing technique is  accurate at the 2\% level over the entire $k$-range of interest.

%% file: results.tex
Several checks were performed to validate our simulations. We first verified  that the power spectrum of  independent 
simulations obtained either with different cosmological parameters or different seeds is consistent with the power spectrum derived from the Taylor expansion of Sec.~\ref{sec:grid}. We then present a comparison of our central model with the  one-dimensional Lyman-$\alpha$ forest power spectrum obtained with BOSS by  \cite{Palanque-Delabrouille2013}.  This allows us to quantify the agreement between our simulation and the measured power spectrum.
Finally, we discuss some characteristics  of our simulations. 
In particular, 
we describe the effect on the
flux power spectrum of some of the  parameters we have varied, and we show the $T-\rho$ diagrams from which we  derive the two parameters $T_0(z)$ and $\gamma(z)$ that describe the IGM.

\subsection{Assessment checks}

We performed three categories of assessment checks. The first one verifies the statistical errors in the simulation relative to the number of lines of sight used to compute the one-dimensional power spectrum. The second one assesses the accuracy of our second-order Taylor expansion to model  the power spectrum  by comparing its prediction to the simulated power spectrum for  simulations  other than those used in the grid. The last category tests the impact of cosmic variance from the use of a given random seed.   

 For each simulation, the one-dimensional power spectrum was computed from 100.000 lines of sight. This large number is  necessary to ensure that the simulation uncertainties remain about an order of magnitude smaller than current most precise data measurements (taken from \citep{Palanque-Delabrouille2013a}). We ensured that the simulation uncertainties were not artificially limited by an oversampling of the simulated volume: this was done by  considering different numbers $n$ of lines of sight (from 5.000 to 100.000).   For each redshift and mode, the power spectrum value  is taken as the mean over the $n$ lines of sight and the uncertainty on the mean as the $rms$ of the distribution divided by $\sqrt{n}$. We checked  that the uncertainty on each point of the power spectrum  scaled as the square-root of the number of lines of sight used to compute it, at better than the percent level. 

To test the accuracy with which    our Taylor expansion reproduces the  power spectrum for different cosmologies, we  performed  simulations with  input cosmological and astrophysical parameters different from those that were used to compute the derivatives. We tested  the most relevant parameters for our study. One simulation was run with different $n_s$ and $\sigma_8$, two others with different $T_0$ and $\gamma$, and a last set with all input parameters different from  their  values in the grid simulations. In each case, we computed the power spectra corresponding to the twelve redshift bins in the range $z= [2.1-4.5]$.  We then performed a simple fit of the six parameters ($n_s$, $\sigma_8$, $\Omega_m$, $H_0$, \texttt{AMPL}, \texttt{GRAD}) using our second-order Taylor expansion as  model. 

The results are summarized in Tab.~\ref{table:val_para}. The last column shows the fitted values over 100.000 lines of sight. The uncertainty  is estimated as the $rms$ of the distribution for each parameter fitted over  10 subsamples of 10.000 lines of sight each, divided by $\sqrt{10}$. The  configurations were chosen so as to probe different relevant regions of the parameter phase space. These tests give results in  excellent agreement with the input  parameters. The level of accuracy achieved with these validation tests is 3 to 5 times better than  the errors  we expect on these parameters from a   fit to data  given the uncertainties of  \citet{Palanque-Delabrouille2013}.  {We note that the accuracy drops rapidly, however, as we test values outside the range that was used to compute the derivatives (cf. table \ref{table:grid_parameters}), as indicated by the almost $3\sigma$ discrepancy on the fitted value  of $T_0$ in the last test. Since our variation range was purposely chosen to be wide enough to include all recent results, this is not expected to cause any problem in the future.} These checks thus demonstrate that our Taylor expansion adequately models the power spectrum for any set of input parameters within the  range of table \ref{table:grid_parameters}.

\begin{table}[h]
\begin{center}
\begin{tabular}{|clcc|}
\hline
Test configuration &Parameter & Input value & Fitted value\\
\hline
\multirow{6}{*}{$n_s - \sigma_8$}&
$n_s$\dotfill & $0.93$ & $0.931\pm\,0.002$\\
&$\sigma_8$\dotfill & $0.85$ & $0.846\pm\,0.008$\\
&$\Omega_m$\dotfill & $0.31$ & $0.310\pm\,0.003$\\
&$H_0$\dotfill & $67.5$ & $67.2\pm\,1.1$\\
&$T_0(z=3)$\dotfill & $14000$ & $14230\pm\,600$\\
&$\gamma(z=3)$\dotfill & $1.32$ & $1.33\pm\,0.03$\\
\hline
\multirow{6}{*}{$T_0 - \gamma$}& % gamma_t0_ter
$n_s$\dotfill & $0.96$ & $0.961\pm\,0.002$\\
&$\sigma_8$\dotfill & $0.83$ & $0.830\pm\,0.009$\\
&$\Omega_m$\dotfill & $0.31$ & $0.310\pm\,0.003$\\
&$H_0$\dotfill & $67.5$ & $67.2\pm\,1.1$\\
&$T_0(z=3)$\dotfill & $10000$ & $10130\pm\,200$\\
&$\gamma(z=3)$\dotfill & $1.47$ & $1.47\pm\,0.02$\\
\hline
\multirow{6}{*}{$T_0 - \gamma$}& % gamma_t0_bis (new)
$n_s$\dotfill & $0.96$ & $0.961\pm\,0.001$\\
&$\sigma_8$\dotfill & $0.83$ & $0.830\pm\,0.008$\\
&$\Omega_m$\dotfill & $0.31$ & $0.310\pm\,0.003$\\
&$H_0$\dotfill & $67.5$ & $67.3\pm\,1.1$\\
&$T_0(z=3)$\dotfill & $10000$ & $10420\pm\,300$\\
&$\gamma(z=3)$\dotfill & $1.16$ & $1.15\pm\,0.02$\\
\hline
\multirow{6}{*}{All parameters}& % all_bis
$n_s$\dotfill & $0.93$ & $0.927\pm\,0.002$\\
&$\sigma_8$\dotfill & $0.86$ & $0.848\pm\,0.004$\\
&$\Omega_m$\dotfill & $0.30$ & $0.300\pm\,0.003$\\
&$H_0$\dotfill & $66$ & $67.7\pm\,1.1$\\
&$T_0(z=3)$\dotfill & $10000$ & $10470\pm\,400$\\
&$\gamma(z=3)$\dotfill & $1.16$ & $1.19\pm\,0.03$\\
\hline
\multirow{6}{*}{All parameters}& % all (new)
$n_s$\dotfill & $0.935$ & $0.935\pm\,0.002$\\
&$\sigma_8$\dotfill & $0.846$ & $0.833\pm\,0.005$\\
&$\Omega_m$\dotfill & $0.285$ & $0.282\pm\,0.004$\\
&$H_0$\dotfill & $68$ & $69.2\pm\,1.1$\\
&$T_0(z=3)$\dotfill & $5840$ & $6720\pm\,320$\\
&$\gamma(z=3)$\dotfill & $1.32$ & $1.30\pm\,0.02$\\
\hline
\end{tabular}
\end{center}
\caption{Comparison of the simulated parameters and the fitted parameters for  different sets of input parameters.}
\label{table:val_para}
\end{table}

\begin{figure}[hh]
\centering
\includegraphics[width=.49\linewidth]{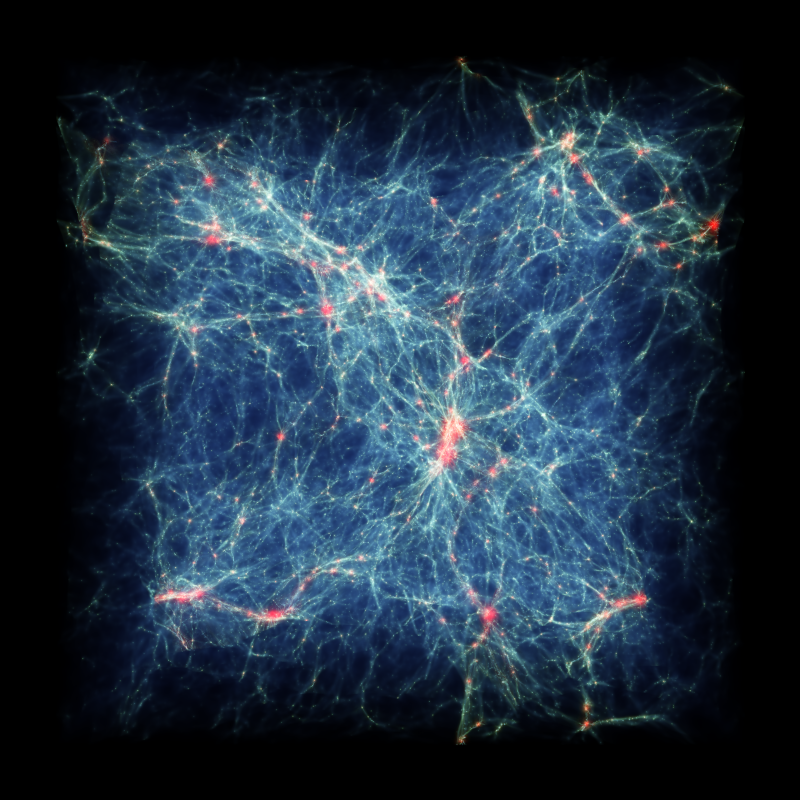}
\includegraphics[width=.49\linewidth]{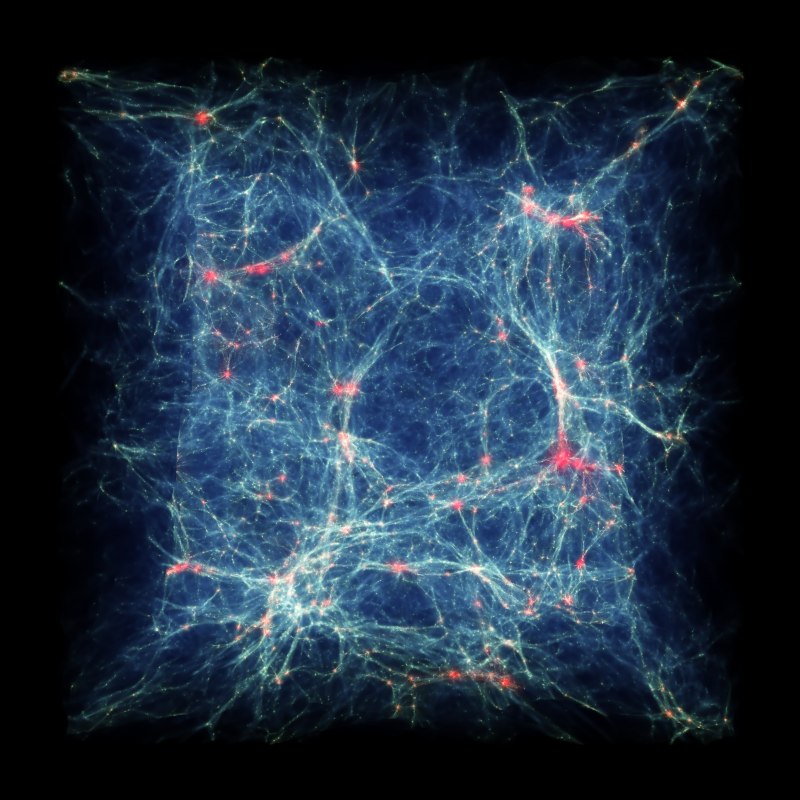}
\caption{Visualisation using \texttt{splotch} of the baryonic gas from a \texttt{GADGET-3} snapshot taken at $z=2.2$ for two simulations run with identical parameters but different random seeds to compute the initial conditions.  Both simulations are using $2x768^3$ particles  in a 
$(\SI{25}{\mega\parsec\per\hubble})^3$ box. Color represents  gas temperature  (from blue to red)
and density is mapped to intensity. Left for the random seed used for the grid, right for a different random seed.}
\label{fig:snapshot}
\end{figure}

Finally, we produced a new simulation with the same parameters as our central simulation but using a different random seed to compute the initial conditions. Snapshots of the resulting gas distribution in the two cases are shown in figure~\ref{fig:snapshot}. The derived power spectra for the two seeds are in excellent agreement on low scales. On the largest scales, the two power spectra can differ by up to 2 to 3$\sigma$ at all redshifts, indicating a sample variance contribution to the uncertainty on the simulated power spectrum due to the fact that the simulation box has a  size close to the largest modes measured. We again performed a simple fit on the power spectrum measured with the new seed using our Taylor expansion as model. The results are given in table~\ref{table:val_seed}. They show that cosmic variance has an impact on the power spectrum that exceeds the simulation statistical uncertainty and will therefore need to be included as a systematic uncertainty when comparing our model to data.  

\begin{table}[h]
\begin{center}
\begin{tabular}{|lcc|}
\hline
Parameter & Input value & Fitted value\\
\hline
$n_s$\dotfill & $0.96$ & $0.969\pm\,0.004$\\
$\sigma_8$\dotfill & $0.83$ & $0.839\pm\,0.005$\\
$\Omega_m$\dotfill & $0.31$ & $0.28\pm\,0.01$\\
%$\Omega_m$\dotfill & $0.31$ & $0.275\pm\,0.010$\\
$H_0$\dotfill & $67.5$ & 65$\pm\,1$\\
$T_0(z=3)$\dotfill & $14000$ & $13750\pm\,1000$\\
$\gamma(z=3)$\dotfill & $1.32$ & $1.38\pm\,0.03$\\
\hline
\end{tabular}
\end{center}
\caption{Comparison of the simulated parameters and the fitted parameters for a different seed in the simulation.}
\label{table:val_seed}
\end{table}

\subsection{Comparison to  SDSS-III/BOSS DR9 data}

\begin{figure}
\centering
\includegraphics[width=\linewidth]{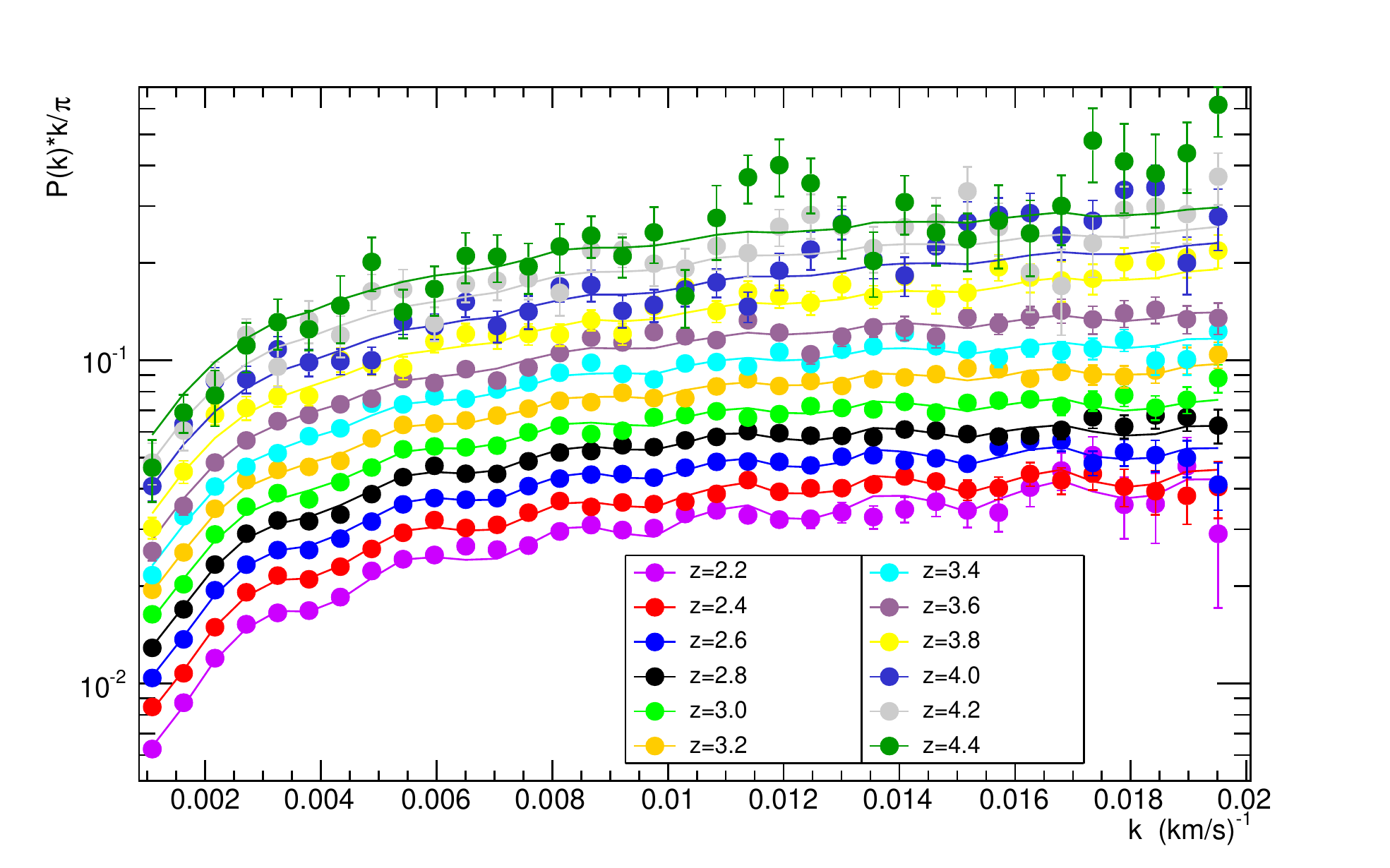}
\caption{One-dimensional Lyman-$\alpha$ forest power spectrum obtained with BOSS spectra. The dots are the measured  power spectrum by \citet{Palanque-Delabrouille2013}). The solid line represents the power spectrum for our central model after adjustment of nuisance parameters to account for imperfect modeling of the instrumental parameters in the 1D power spectrum measurement.}
\label{fig:datafit}
\end{figure}

In \citet{Palanque-Delabrouille2013}, the one-dimensional Lyman-$\alpha$ forest power spectrum is measured with  13,821 quasar spectra from SDSS-III/BOSS DR9  selected  on the basis of their high quality,  large signal-to-noise ratio, and  good spectral resolution. The  power spectra  are measured over twelve redshift bins from $\langle z\rangle = 2.2$ to $\langle z\rangle = 4.4$, and  scales from 0.001~$\rm(km/s)^{-1}$ to $0.02~\rm(km/s)^{-1}$ (see figure~\ref{fig:datafit}). 

In order to compare the measurements to the power spectrum obtained for our central model model, we normalized the simulation power spectrum at each redshift by constraining the effective optical depth to follow the power law evolution $\tau_{\rm eff}(z) =  \tau_A\times (1+z) ^{\tau_{S}}$, where $\tau_A=0.0025$ and $\tau_{S}=3.7$. To account for the effect of the correlated \ion{Si}{iii}  absorption, we  correct the simulated power spectrum by a multiplicative  term, $1+a^2+2a\cos(vk)$ with $a = f_ {\rm{Si\,III}}/(1-{ \left< F \right>}(z))$ following the suggestion  of~\citet{McDonald2006}. The parameter $f_ {\rm{Si\,III}}$ is adjusted and $v$ is fixed at 2271~km/s. We  model the imperfection of the resolution of BOSS spectra though a multiplicative term. Finally, we allow for imperfection in the noise estimate of the BOSS spectra with eight additive terms (one for each redshift bin). 

Figure \ref{fig:datafit} illustrates the good agreement between the data and the simulations. Without any adjustment of the cosmological and astrophysical parameters, the $\chi^2$ per number of degrees of freedom is already better than 1.2. The good agreement between data and simulation covers the whole redshift range,  $z= [2.1-4.5]$, in contrast with the cosmological analysis described in  \citet{Palanque-Delabrouille2013} which was performed over the reduced redshift range $z= [2.1-3.7]$. This simple comparison demonstrates the improvement obtained with these simulations over the previous generation of simulations  \citep{ Viel2010}.

\subsection{Power spectrum}
\begin{figure*}
\centering
\includegraphics[width=\linewidth]{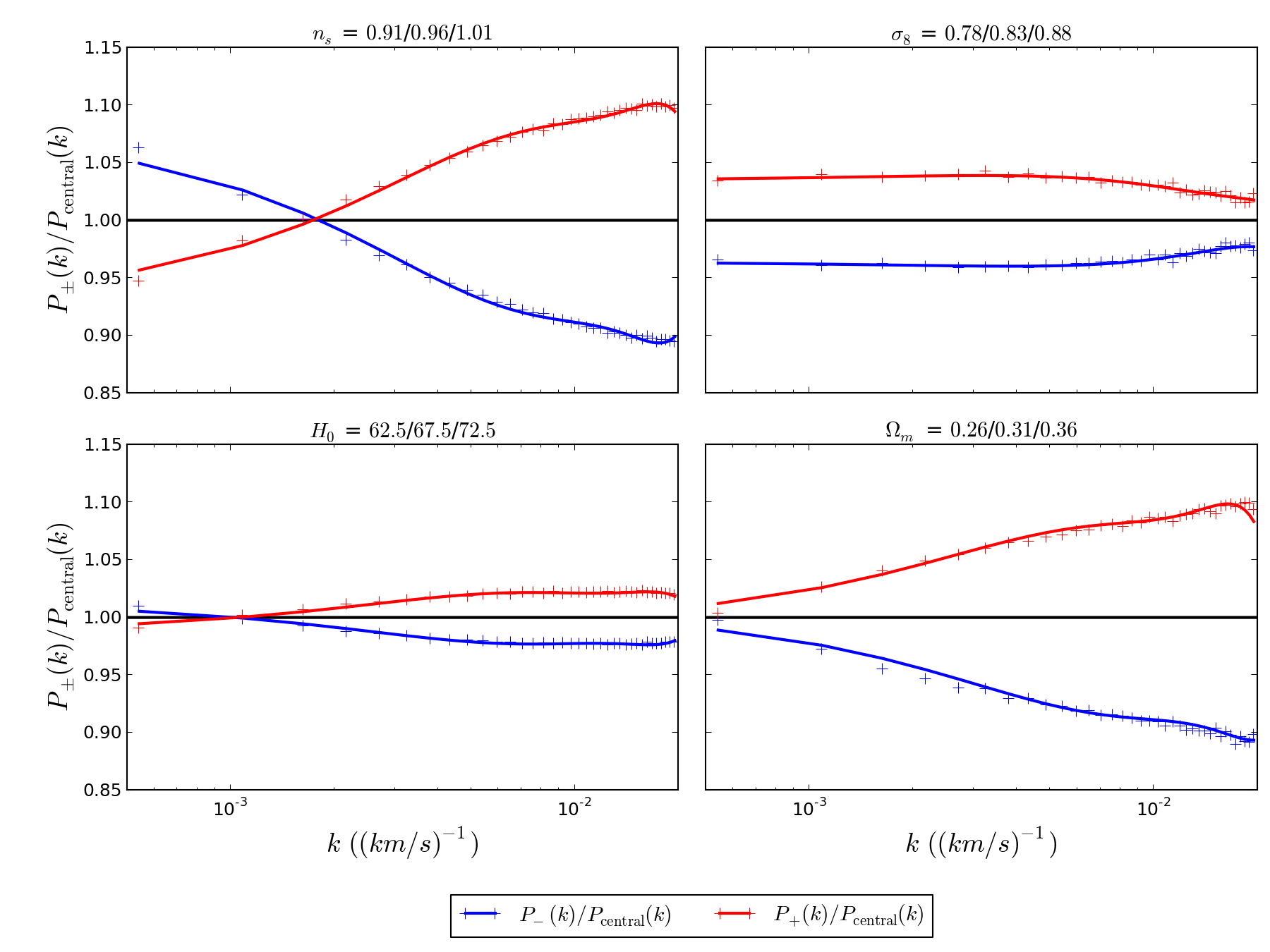}
\caption{Effect of the parameters $n_s$, $\sigma_8$, $H_0$ and $\Omega_m$ on the power spectrum (limited to the $k$-range of our
study) at $z=3.2$. $P_+(k)$ and $P_-(k)$ refer to the power spectra extracted from the simulations using the upper and lower limit on each considered parameter respectively. The fit to the points is a 5\textsuperscript{th} order polynomial function.}\label{fig:parameters_effect}
\end{figure*}

Figure \ref{fig:parameters_effect} illustrates the impact on the power spectrum of our four cosmological parameters. We compare the power spectrum computed from our best-guess model to the one obtained when varying each parameter, one at a time. We note that the dependence on the value of the four parameters is as expected according to their physical meaning. We briefly explain the different behaviors below. 

The spectral index $n_s$ represents the evolution of the primordial density fluctuations with respect to $k$ through
$\mathcal{P}(k) \propto k^{n_s -1}$.  A larger $n_s$ therefore increases the power at large $k$, as seen in the top left panel of figure~\ref{fig:parameters_effect}.

The parameter {$\sigma_8$} measures the $rms$ amplitude of the linear matter density fluctuations today in spheres  of size $8\,h^{-1}\,{\rm Mpc}$, and thus determines the normalization of the matter power spectrum.  To first order, increasing the value of  {$\sigma_8$} therefore  increases
the power spectrum on all scales, as shown in the top right panel of figure~\ref{fig:parameters_effect}. A slightly larger effect, however, is  seen on large scales, since an excess in the amplitude of the fluctuations will favor the merging of small scale fluctuations, thus enhancing the power on larger scales. This tiny trend is purely non-linear and not expected in the evolution with $\sigma_8$ of the linear power-spectrum. 

The present-day Hubble constant {$H_0$} (in units of velocity/distance) allows the conversion from distance-space to $k$-space (units of inverse velocity). Therefore,  if $H_0$ is increased, a given distance will correspond to a higher $k$, thus leading to an increase of power since the power spectrum, which is a decreasing function of $k$, 
is shifted to the right. This is indeed what is observed in the lower left panel of figure~\ref{fig:parameters_effect}.

Finally, the parameter {$\Omega_m$} quantifies the fraction of matter density in a flat Universe. Because $\Omega_m$ and  the dark energy density $\Omega_\Lambda$ vary in  opposite directions, a
higher $\Omega_m$ delays the onset of dark energy domination, thus increasing the time available for structure formation.
In addition, in a larger $\Omega_m$ universe, more structures (in particular  small ones that would not collapse otherwise) will be formed, 
leading to an increase of the power spectrum,  especially at high $k$. This is in agreement with the plots in the lower right panel of figure~\ref{fig:parameters_effect}.

\subsection{Density-temperature relation}
\begin{figure*}
\centering
\includegraphics[width=\linewidth]{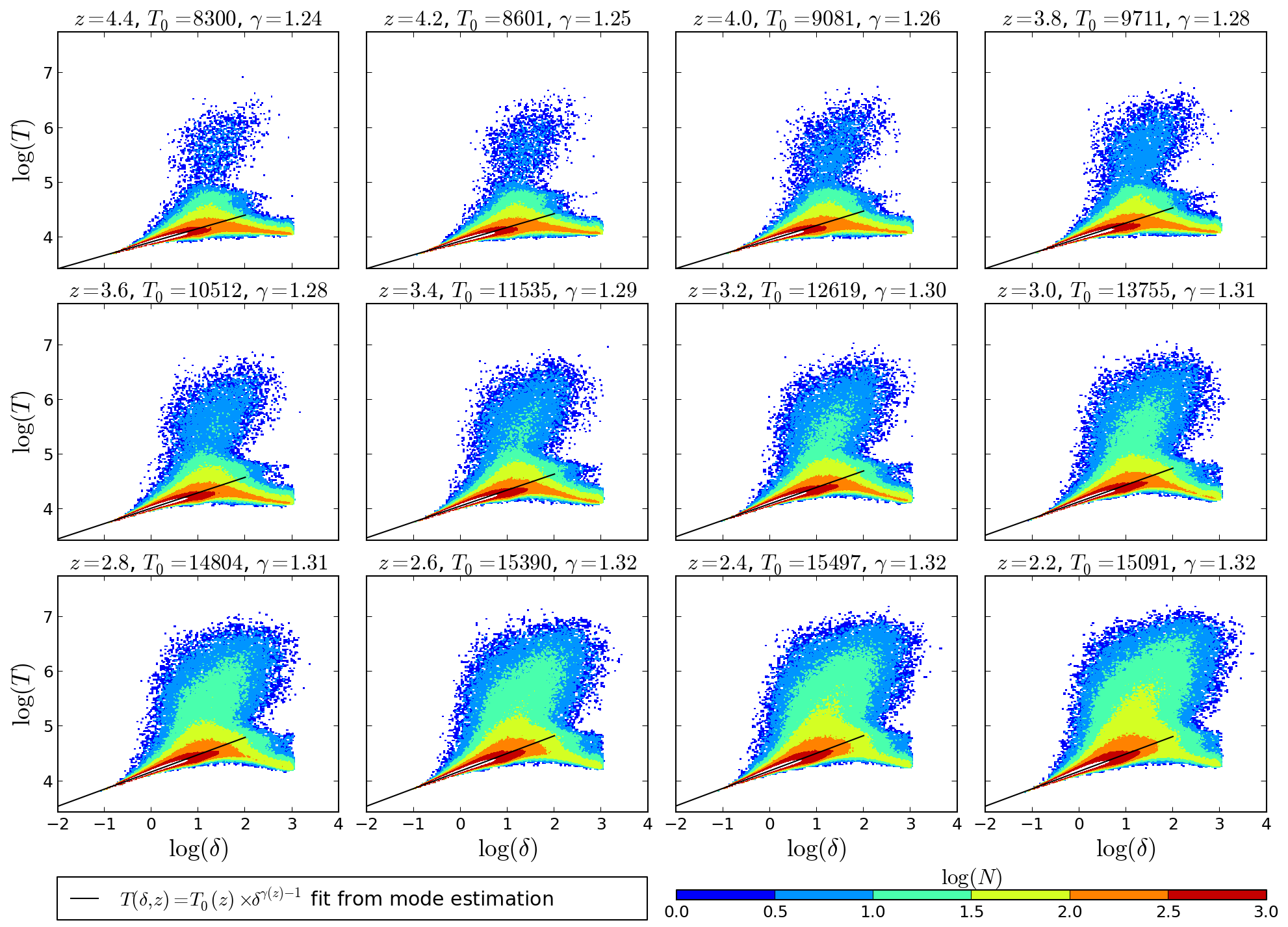}
\caption{Temperature-density diagrams at various redshift. Color represents the particle density in logarithmic scale. The black line represents the
fitted $T-\rho$ relation from several mode-estimated points. $\delta$ is the normalized density $\rho / \left< \rho \right>$.}\label{fig:t-rho}
\end{figure*}

In figure~\ref{fig:t-rho}, we present the $T-\rho$ diagrams obtained from our central simulation at each of the snapshot redshifts. We can distinguish three different populations -- the IGM, the stars, and the clusters -- with a clear evolution with redshift for each of them.  

\begin{figure*}
\centering
\includegraphics[width=\linewidth]{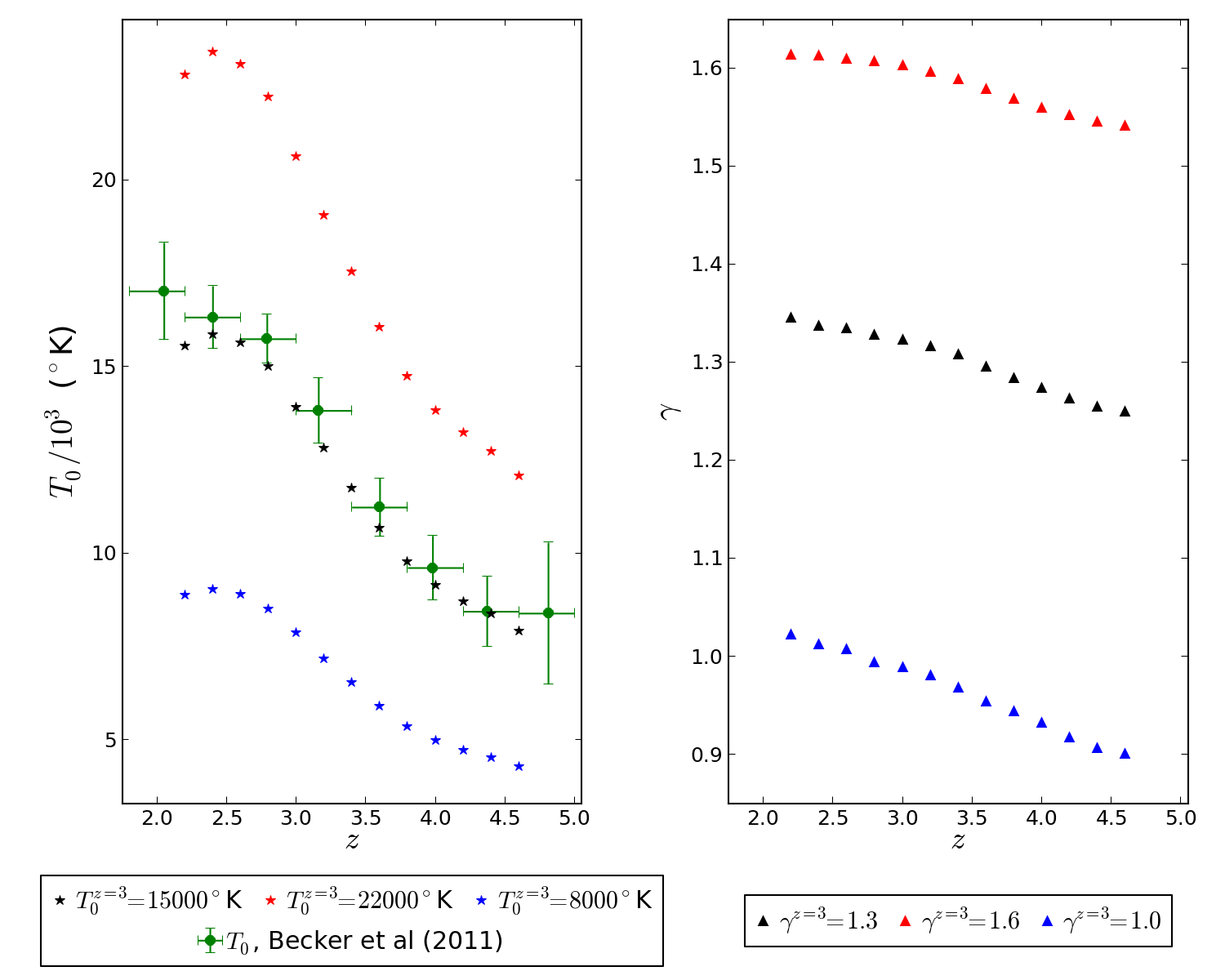}
\caption{Evolution with redshift of $T_0$ and $\gamma$  for the different values of these two parameters used in the grid. The plotted values are extracted from a
sample of particles using a mode estimation as explained in the text. We overlaid the measurements of \citet{Becker2011} (for $\gamma=1.3$) for comparison.}\label{fig:t0_gamma}
\end{figure*}

The IGM  is described by the low density and low temperature particles. This is the regime that dominates at high redshift. At later times, however, fewer and fewer particles reside in this part of the $T-\rho$ diagram, since they are captured by collapsing over-densities. We use this region  to extract the 
$T_0(z)$ and $\gamma(z)$ parameters,   displayed in figure \ref{fig:t0_gamma}, where they are compared to the measurements of \citet{Becker2011}.

The particles with  higher temperature correspond to clusters and galactic gas. As expected, their density increases as structures are formed in the simulation box. They therefore become more prominent at lower redshifts.

In our simulations dedicated to the study of the IGM though the Lyman-$\alpha$ forest measurements, star formation undergoes a simplified treatment, which  reflects as  the sharp cut-off at $\log(\delta) \simeq 3$. Any particle sufficiently
dense and cool is transformed into a star particle. The latter is used for gravity force calculation, but  does not undergo SPH treatment like baryonic gas does.

%% file: conclusion.tex
We have designed and produced a grid of cosmological simulations, which may be used to extract constraints on cosmological
parameters from Lyman-$\alpha$ surveys, whether   current like SDSS-III/BOSS or  future like SDSS-IV/eBOSS.
These simulations cover the redshift range $2.2-4.6$. They explore the cosmological parameters $n_s$, $\sigma_8$, $H_0$, and $\Omega_m$ over a large range centered on Planck measurements, as well as the astrophysical parameters 
$T_0$ and $\gamma$ in a range covering most  recent results. 

Using the splicing technique of \citet{McDonald2003}, we computed 
1D power spectra from simulations equivalent to a $\SI{100}{\mega\parsec\per\hubble}$ box filled
with $3072^2$ particles of each species (here dark matter and baryonic gas), abbreviated to (100, 3072) using our standard 
notation, from lower-resolution (100,768) and smaller box-size (25,768) simulations, combined using a transition (25,192) simulation. We show that the splicing technique allows us to approximate the exact full-resolution large-box simulation with an accuracy at the 2\% level. 

While one full-size high-resolution (100,3072) simulation would have required of order one million hours, one equivalent set of 3 simulations consumes an average of 70,000 hours of CPU time, with most of the time in the 
simulation pipeline (see  figure ~\ref{fig:pipeline}) being spent on performing the hydrodynamical 
simulations. The data volume produced by each set is 1.6 terabytes. Therefore, the whole grid represents about
2 millions hours of CPU time and a volume of 45 terabytes of data.

From the 1D power-spectra that we computed at each point of the grid, we derived a second-order Taylor expansion around our best-guess model. It describes the evolution of the 1D power spectrum with changes in either the cosmological or the astrophysical parameters that we studied. We have
performed several  check runs to ensure the quality and validity of our simulation grid, using either different seeds, or off-the-grid values of the cosmological and astrophysical parameters. These checks were all consistent with the power spectrum predicted using our second-order Taylor expansion, thus validating it. We compared our central simulation  to published data from BOSS and showed that they were already in good agreement without any adjustment of any of the simulation  parameters. 
In forthcoming work, we will use this Taylor expansion for a  quantitative
comparison to  data in order to extract best-fit cosmological parameters. 

These simulations are  accompanied by a set of simulations  where massive neutrinos are included. These required 
 additional  developments for an efficient treatment and a proper account of the additional particles (at all levels of the pipeline: in {\ttfamily CAMB}, in the setup of the initial conditions for thermal velocities, in {\ttfamily Gadget-3}, etc.), but are otherwise produced with a pipeline similar to  the one presented in this study. The details about the simulations with massive neutrinos can be found in the companion paper~\citep{Rossi2014}. 
Additional parameters can yet be included in the same context.   However, due to the
presence of the cross terms that are necessary for an accurate modeling of the likelihood function that illustrates the variation of the power spectrum in all directions of this growing parameter-space, adding new parameters will become more and more expensive in terms of calculation time.

%% file: appendix.tex
We  summarize in table~\ref{tab:simsgrid} and \ref{tab:simstests} all the simulations mentioned in the paper. For box size and number of particles, $(L,N)$ refers to a simulation with $N^3$ particles per species (gas or dark matter, thus $2\times N^3$ particles total) in a box of size 
$L\:\SI{}{\mega\parsec\per\hubble}$ on a side. Standard parameters are $(n_s, \sigma_8,\Omega_m,H_0,T_0,\gamma)= (0.83,0.96,0.31,67.5,14.000,1.3)$. Unless parameter names are explicitly listed, values are given for all parameters in the order just mentioned. All parameters  are assumed to have their standard value unless specified otherwise. Except for the simulations performed for the convergence tests or to compute the exact power spectrum in the splicing test, all simulations are using the splicing technique to combine each set of three simulations into a single one of equivalent size to the largest box and equivalent mass-resolution to the best mass resolution. In the first column of the table \ref{tab:simsgrid}, $\partial_i$ and $\partial_{ii}$ indicate simulations needed to compute the first and second order derivatives with respect to parameter $i$,  and $\partial_{ij}$ indicate additional simulations required to compute cross-derivatives with respect to parameters $i$ and $j$, where  $i$ and $j$ are explicitly given in the same column.  In the last column of the same table,
we also give the values of the shape parameter $\Gamma= \Omega_m h$ and of $\sigma_8(z=3.0)$ where $z=3.0$ corresponds to the pivot redshift of  Ly$\alpha$ power spectrum measurements  in  \cite{Palanque-Delabrouille2013} and \cite{McDonald2003}. These parameters are not used  in this work but we make them available since they  are closely related to observations and  were often used in the pioneering work on  Ly$\alpha$ forest measurements.

\begin{longtable}{|l|c|c|c|}
\hline
\multicolumn{4}{|c|}{\bf Grid}\\
\hline
Type& Box size,  particles $(L,N)$ & Simulation parameters &$\sigma_8(z=3)$; $\Gamma$\\
\hline
Central & (25,768)+(100,768)+(25,192) & Standard &0.26; 0.21\\
   $\partial_{i}$, $\partial_{ii}:$ $n_s$ & (25,768)+(100,768)+(25,192) & $n_s=0.91$& 0.26; 0.21 \\
 $\partial_{i}$, $\partial_{ii}:$ $n_s$ & (25,768)+(100,768)+(25,192) & $n_s=1.01$ &0.26; 0.21\\
  $\partial_{i}$, $\partial_{ii}:$ $\sigma_8$ & (25,768)+(100,768)+(25,192) & $\sigma_8=0.83$&0.25; 0.21\\
  $\partial_{i}$, $\partial_{ii}:$ $\sigma_8$ & (25,768)+(100,768)+(25,192) & $\sigma_8=0.93$&0.28; 0.21\\
 $\partial_{i}$, $\partial_{ii}:$ $\Omega_m$ & (25,768)+(100,768)+(25,192) & $\Omega_m=0.26$& 0.27; 0.18\\
 $\partial_{i}$, $\partial_{ii}:$ $\Omega_m$ & (25,768)+(100,768)+(25,192) &  $\Omega_m=0.36$ &0.25; 0.24\\
 $\partial_{i}$, $\partial_{ii}:$ $H_0$ & (25,768)+(100,768)+(25,192) & $H_0=62.5$ &0.26; 0.19\\
  $\partial_{i}$, $\partial_{ii}:$  $H_0$ & (25,768)+(100,768)+(25,192) & $H_0=72.5$ &0.26; 0.22\\
  $\partial_{i}$, $\partial_{ii}:$  $T_0$ & (25,768)+(100,768)+(25,192) & $T_0=7.000$& 0.26; 0.21\\
  $\partial_{i}$, $\partial_{ii}:$  $T_0$ & (25,768)+(100,768)+(25,192) & $T_0=21.000$& 0.26; 0.21\\
  $\partial_{i}$, $\partial_{ii}:$  $\gamma$ & (25,768)+(100,768)+(25,192) & $\gamma=1.0$&0.26; 0.21\\
  $\partial_{i}$, $\partial_{ii}:$  $\gamma$ & (25,768)+(100,768)+(25,192) &  $\gamma=1.6$ &0.26; 0.21\\
  $\partial_{ij}:$ $n_s - \sigma_8$ & (25,768)+(100,768)+(25,192) & $n_s=1.01$, $\sigma_8=0.93$& 0.28; 0.21\\
  $\partial_{ij}:$ $n_s - \Omega_m$ & (25,768)+(100,768)+(25,192) & $n_s=1.01$, $\Omega_m=0.36$ &0.25; 0.24\\
  $\partial_{ij}:$ $n_s - H_0$ & (25,768)+(100,768)+(25,192) & $n_s=1.01$, $H_0=72.5$ &0.26; 0.22\\
  $\partial_{ij}:$ $n_s - T_0$ & (25,768)+(100,768)+(25,192) & $n_s=1.01$, $T_0=21.000$&0.26; 0.21\\
  $\partial_{ij}:$ $n_s -\gamma$ & (25,768)+(100,768)+(25,192) & $n_s=1.01$, $\gamma=1.6$&0.26; 0.21\\
  $\partial_{ij}:$ $\sigma_8-\Omega_m$& (25,768)+(100,768)+(25,192) &  $\sigma_8=0.93$, $\Omega_m=0.36$&0.27; 0.24\\
  $\partial_{ij}:$  $\sigma_8-H_0$& (25,768)+(100,768)+(25,192) &  $\sigma_8=0.93$, $H_0=72.5$&0.28; 0.22\\
  $\partial_{ij}:$  $\sigma_8-T_0$& (25,768)+(100,768)+(25,192) &  $\sigma_8=0.93$, $T_0=21.000$&0.28; 0.21\\
  $\partial_{ij}:$  $\sigma_8-\gamma$& (25,768)+(100,768)+(25,192) &  $\sigma_8=0.93$, $\gamma=1.6$&0.28; 0.21\\
  $\partial_{ij}:$  $\Omega_m-H_0$& (25,768)+(100,768)+(25,192) &  $\Omega_m=0.36$, $H_0=72.5$&0.25; 0.26\\
  $\partial_{ij}:$  $\Omega_m-T_0$& (25,768)+(100,768)+(25,192) &  $\Omega_m=0.36$, $T_0=21.000$&0.25; 0.24\\
  $\partial_{ij}:$  $\Omega_m-\gamma$& (25,768)+(100,768)+(25,192) &  $\Omega_m=0.36$, $\gamma=1.6$&0.25; 0.24\\
  $\partial_{ij}:$  $H_0-T_0$& (25,768)+(100,768)+(25,192) & $H_0=72.5$, $T_0=21.000$ &0.26; 0.22\\
  $\partial_{ij}:$  $H_0-\gamma$& (25,768)+(100,768)+(25,192) &  $H_0=72.5$, $\gamma=1.6$&0.26; 0.22\\
  $\partial_{ij}:$  $T_0-\gamma$& (25,768)+(100,768)+(25,192) & $T_0=21.000$, $\gamma=1.6$ &0.26; 0.21\\
\hline

\caption{Simulations used in this work for the grid. }
\label{tab:simsgrid}
\end{longtable}

\begin{longtable}{|l|c|c|}
\hline
\multicolumn{3}{|c|}{\bf Convergence tests}\\
\hline
Type& Box size,  particles $(L,N)$ & Simulation parameters \\
\hline
 Resolution & (20,1024) & Standard \\
Resolution & (20,768) & Standard \\
Resolution & (20,512) & Standard \\
Resolution & (20,384) & Standard \\
Resolution & (20,192) & Standard \\
Box size & (120,1024) & Standard \\
Box size & (90,768) & Standard \\
Box size & (80,683) & Standard \\
Box size & (60,512) & Standard \\
Box size & (20,171) & Standard \\
\hline
\multicolumn{3}{|c|}{\bf Splicing tests}\\
\hline
Type& Box size,  particles $(L,N)$ & Simulation parameters \\
\hline
Grid-like  & (25,256)+(100,256)+(25,64) & Standard\\
Exact  & (100,1024) & Standard \\
\hline
\multicolumn{3}{|c|}{\bf Validity checks}\\
\hline
Type& Box size,  particles $(L,N)$ & Simulation parameters \\
\hline
Random seed & (25,768)+(100,768)+(25,192) & Standard \\
Off-grid 1 & (25,768)+(100,768)+(25,192) & $(0.93,0.85,0.31,67.5,14.000,1.32)$ \\
Off-grid 2 & (25,768)+(100,768)+(25,192) & $(0.96,0.83,0.31,67.5,10.000,1.47)$ \\
Off-grid 3 & (25,768)+(100,768)+(25,192) & $(0.96,0.83,0.31,67.5,10.000,1.16)$ \\
 Off-grid 4 & (25,768)+(100,768)+(25,192) & $(0.93,0.86,0.30,66,10.000,1.16)$ \\
 Off-grid 5 & (25,768)+(100,768)+(25,192) & $(0.935,0.846,0.285,68,5.840,1.32)$ \\
\hline

\caption{Simulations used in this work for the tests and final validity checks. }
 \label{tab:simstests}
\end{longtable}